\documentclass[12pt]{article}
\usepackage[a4paper, left=2.5cm, right=2.5cm, top=3cm, bottom=3cm]{geometry}
\usepackage{amsmath, amssymb, comment, accents, slashed, sectsty}
\numberwithin{equation}{section}
\sectionfont{\large}
\subsectionfont{\normalsize}

\newcommand{\del}{\partial}
\newcommand{\delb}{{\bar\partial}}

\newcommand{\ket}[1]{|#1\rangle}

\newcommand{\vev}[1]{\langle #1 \rangle}
\newcommand{\bigvev}[1]{\bigl\langle #1 \bigr\rangle}
\newcommand{\Bigvev}[1]{\Bigl\langle #1 \Bigr\rangle}

\newcommand{\ch}{\mathop{\mathrm{ch}}\nolimits}

\newcommand{\rank}{\mathop{\mathrm{rank}}\nolimits}

\newcommand{\SL}{\mathrm{SL}}

\newcommand{\PGL}{\mathrm{PGL}}
\newcommand{\U}{\mathrm{U}}

\newcommand{\longto}{\longrightarrow}
\newcommand{\iso}{\cong}

\newcommand{\Z}{\mathbb{Z}}

\newcommand{\R}{\mathbb{R}}
\newcommand{\C}{\mathbb{C}}

\newcommand{\CP}{\mathbb{CP}}

\let\nc\newcommand
\let\renc\renewcommand
\nc{\wbar}{\overline}
\let\td\tilde
\let\wtd\widetilde
\let\wht\widehat
\let\mcl\mathcal

\nc{\ab}{{\bar{a}}} \nc{\at}{\tilde{a}} \nc{\ah}{\hat{a}}
\nc{\bb}{{\bar{b}}} \nc{\bt}{\tilde{b}} \nc{\bh}{\hat{b}}
\nc{\cb}{{\bar{c}}} \nc{\ct}{\tilde{c}} 
\nc{\db}{{\bar{d}}} \nc{\dt}{\tilde{d}} \renc{\dh}{\hat{d}}
\nc{\eb}{{\bar{e}}} \nc{\et}{\tilde{e}} \nc{\eh}{\hat{e}}
\nc{\fb}{{\bar{f}}} \nc{\ft}{\tilde{f}} \nc{\fh}{\hat{f}}
\nc{\gb}{{\bar{g}}} \nc{\gt}{\tilde{g}} \nc{\gh}{\hat{g}}
\nc{\hb}{{\bar{h}}} \nc{\hh}{\hat{h}} 
\nc{\ib}{{\bar{\imath}}} \nc{\ih}{\hat{\imath}} 
\nc{\jb}{{\bar{\jmath}}} \nc{\jt}{\tilde{\jmath}} \nc{\jh}{\hat{\jmath}}
\nc{\kb}{{\bar{k}}} \nc{\kt}{\tilde{k}} \nc{\kh}{\hat{k}}
\nc{\lb}{{\bar{l}}} \nc{\lt}{\tilde{l}} \nc{\lh}{\hat{l}}
\nc{\mb}{{\bar{m}}} \nc{\mt}{\tilde{m}} \nc{\mh}{\hat{m}}
\nc{\nb}{{\bar{n}}} \nc{\nt}{\tilde{n}} \nc{\nh}{\hat{n}}
\nc{\ob}{{\bar{o}}} \nc{\ot}{\tilde{o}} \nc{\oh}{\hat{o}}
\nc{\pb}{{\bar{p}}} \nc{\pt}{\tilde{p}} \nc{\ph}{\hat{p}}
\nc{\qb}{{\bar{q}}} \nc{\qt}{\tilde{q}} \nc{\qh}{\hat{q}}
\nc{\rb}{{\bar{r}}} \nc{\rt}{\tilde{r}} \nc{\rh}{\hat{r}}
\renc{\sb}{{\bar{s}}} \nc{\st}{\tilde{s}} \nc{\sh}{\hat{s}}
\nc{\tb}{{\bar{t}}} \renc{\th}{\hat{t}} 
\nc{\ub}{{\bar{u}}} \nc{\ut}{\tilde{u}} \nc{\uh}{\hat{u}}
\nc{\vb}{{\bar{v}}} \nc{\vt}{\tilde{v}} \nc{\vh}{\hat{v}}
\nc{\wb}{{\bar{w}}} \nc{\wt}{\tilde{w}} \nc{\wh}{\hat{w}}
\nc{\xb}{{\bar{x}}} \nc{\xt}{\tilde{x}} \nc{\xh}{\hat{x}}
\nc{\yb}{{\bar{y}}} \nc{\yt}{\tilde{y}} \nc{\yh}{\hat{y}}
\nc{\zb}{{\bar{z}}} \nc{\zt}{\tilde{z}} \nc{\zh}{\hat{z}}

\nc{\Ab}{\wbar{A}} \nc{\At}{\wtd{A}} \nc{\Ah}{\wht{A}}
\nc{\Bb}{\wbar{B}} \nc{\Bt}{\wtd{B}} \nc{\Bh}{\wht{B}}
\nc{\Cb}{\wbar{C}} \nc{\Ct}{\wtd{C}} \nc{\Ch}{\wht{C}}
\nc{\Db}{\wbar{D}} \nc{\Dt}{\wtd{D}} \nc{\Dh}{\wht{D}}
\nc{\Eb}{\wbar{E}} \nc{\Et}{\wtd{E}} \nc{\Eh}{\wht{E}}
\nc{\Fb}{\wbar{F}} \nc{\Ft}{\wtd{F}} \nc{\Fh}{\wht{F}}
\nc{\Gb}{\wbar{G}} \nc{\Gt}{\wtd{G}} \nc{\Gh}{\wht{G}}
\nc{\Hb}{\wbar{H}} \nc{\Ht}{\wtd{H}} \nc{\Hh}{\wht{H}}
\nc{\Ib}{\wbar{I}} \nc{\It}{\wtd{I}} \nc{\Ih}{\wht{I}}
\nc{\Jb}{\wbar{J}} \nc{\Jt}{\wtd{J}} \nc{\Jh}{\wht{J}}
\nc{\Kb}{\wbar{K}} \nc{\Kt}{\wtd{K}} \nc{\Kh}{\wht{K}}
\nc{\Lb}{\wbar{L}} \nc{\Lt}{\wtd{L}} \nc{\Lh}{\wht{L}}
\nc{\Mb}{\wbar{M}} \nc{\Mt}{\wtd{M}} \nc{\Mh}{\wht{M}}
\nc{\Nb}{\wbar{N}} \nc{\Nt}{\wtd{N}} \nc{\Nh}{\wht{N}}
\nc{\Ob}{\wbar{O}} \nc{\Ot}{\wtd{O}} \nc{\Oh}{\wht{O}}
\nc{\Pb}{\wbar{P}} \nc{\Pt}{\wtd{P}} \nc{\Ph}{\wht{P}}
\nc{\Qb}{\wbar{Q}} \nc{\Qt}{\wtd{Q}} \nc{\Qh}{\wht{Q}}
\nc{\Rb}{\wbar{R}} \nc{\Rt}{\wtd{R}} \nc{\Rh}{\wht{R}}
\nc{\Sb}{\wbar{S}} \nc{\St}{\wtd{S}} \nc{\Sh}{\wht{S}}
\nc{\Tb}{\wbar{T}} \nc{\Tt}{\wtd{T}} \nc{\Th}{\wht{T}}
\nc{\Ub}{\wbar{U}} \nc{\Ut}{\wtd{U}} \nc{\Uh}{\wht{U}}
\nc{\Vb}{\wbar{V}} \nc{\Vt}{\wtd{V}} \nc{\Vh}{\wht{V}}
\nc{\Wb}{\wbar{W}} \nc{\Wt}{\wtd{W}} \nc{\Wh}{\wht{W}}
\nc{\Xb}{\wbar{X}} \nc{\Xt}{\wtd{X}} \nc{\Xh}{\wht{X}}
\nc{\Yb}{\wbar{Y}} \nc{\Yt}{\wtd{Y}} \nc{\Yh}{\wht{Y}}
\nc{\Zb}{\wbar{Z}} \nc{\Zt}{\wtd{Z}} \nc{\Zh}{\wht{Z}}

\nc{\CA}{\mcl{A}} \nc{\CAb}{\wbar{\CA}} \nc{\CAt}{\wtd{\CA}} \nc{\CAh}{\wht{\CA}}
\nc{\CB}{\mcl{B}} \nc{\CBb}{\wbar{\CB}} \nc{\CBt}{\wtd{\CB}} \nc{\CBh}{\wht{\CB}}
\nc{\CC}{\mcl{C}} \nc{\CCb}{\wbar{\CC}} \nc{\CCt}{\wtd{\CC}} \nc{\CCh}{\wht{\CC}}
\nc{\CD}{\mcl{D}} \nc{\CDb}{\wbar{\CD}} \nc{\CDt}{\wtd{\CD}} \nc{\CDh}{\wht{\CD}}
\nc{\CE}{\mcl{E}} \nc{\CEb}{\wbar{\CE}} \nc{\CEt}{\wtd{\CE}} \nc{\CEh}{\wht{\CE}}
\nc{\CF}{\mcl{F}} \nc{\CFb}{\wbar{\CF}} \nc{\CFt}{\wtd{\CF}} \nc{\CFh}{\wht{\CF}}
\nc{\CG}{\mcl{G}} \nc{\CGb}{\wbar{\CG}} \nc{\CGt}{\wtd{\CG}} \nc{\CGh}{\wht{\CG}}
\nc{\CH}{\mcl{H}} \nc{\CHb}{\wbar{\CH}} \nc{\CHt}{\wtd{\CH}} \nc{\CHh}{\wht{\CH}}
\nc{\CI}{\mcl{I}} \nc{\CIb}{\wbar{\CI}} \nc{\CIt}{\wtd{\CI}} \nc{\CIh}{\wht{\CI}}
\nc{\CJ}{\mcl{J}} \nc{\CJb}{\wbar{\CJ}} \nc{\CJt}{\wtd{\CJ}} \nc{\CJh}{\wht{\CJ}}
\nc{\CK}{\mcl{K}} \nc{\CKb}{\wbar{\CK}} \nc{\CKt}{\wtd{\CK}} \nc{\CKh}{\wht{\CK}}
\nc{\CL}{\mcl{L}} \nc{\CLb}{\wbar{\CL}} \nc{\CLt}{\wtd{\CL}} \nc{\CLh}{\wht{\CL}}
\nc{\CM}{\mcl{M}} \nc{\CMb}{\wbar{\CM}} \nc{\CMt}{\wtd{\CM}} \nc{\CMh}{\wht{\CM}}
\nc{\CN}{\mcl{N}} \nc{\CNb}{\wbar{\CN}} \nc{\CNt}{\wtd{\CN}} \nc{\CNh}{\wht{\CN}}
\nc{\CO}{\mcl{O}} \nc{\COb}{\wbar{\CO}} \nc{\COt}{\wtd{\CO}} \nc{\COh}{\wht{\CO}}
\nc{\CQ}{\mcl{Q}} \nc{\CQb}{\wbar{\CQ}} \nc{\CQt}{\wtd{\CQ}} \nc{\CQh}{\wht{\CQ}}
\nc{\CR}{\mcl{R}} \nc{\CRb}{\wbar{\CR}} \nc{\CRt}{\wtd{\CR}} \nc{\CRh}{\wht{\CR}}
\nc{\CS}{\mcl{S}} \nc{\CSb}{\wbar{\CS}} \nc{\CSt}{\wtd{\CS}} \nc{\CSh}{\wht{\CS}}
\nc{\CT}{\mcl{T}} \nc{\CTb}{\wbar{\CT}} \nc{\CTt}{\wtd{\CT}} \nc{\CTh}{\wht{\CT}}
\nc{\CU}{\mcl{U}} \nc{\CUb}{\wbar{\CU}} \nc{\CUt}{\wtd{\CU}} \nc{\CUh}{\wht{\CU}}
\nc{\CV}{\mcl{V}} \nc{\CVb}{\wbar{\CV}} \nc{\CVt}{\wtd{\CV}} \nc{\CVh}{\wht{\CV}}
\nc{\CW}{\mcl{W}} \nc{\CWb}{\wbar{\CW}} \nc{\CWt}{\wtd{\CW}} \nc{\CWh}{\wht{\CW}}
\nc{\CX}{\mcl{X}} \nc{\CXb}{\wbar{\CX}} \nc{\CXt}{\wtd{\CX}} \nc{\CXh}{\wht{\CX}}
\nc{\CY}{\mcl{Y}} \nc{\CYb}{\wbar{\CY}} \nc{\CYt}{\wtd{\CY}} \nc{\CYh}{\wht{\CY}}
\nc{\CZ}{\mcl{Z}} \nc{\CZb}{\wbar{\CZ}} \nc{\CZt}{\wtd{\CZ}} \nc{\CZh}{\wht{\CZ}}

\let\eps\epsilon
\let\ups\upsilon
\let\veps\varepsilon
\let\vtht\vartheta
\let\vsgm\varsigma
\let\vphi\varphi
\let\vrho\varrho

\nc{\alphab}{\bar{\alpha}} \nc{\alphat}{\td{\alpha}} \nc{\alphah}{\hat{\alpha}}
\nc{\betab}{\bar{\beta}}   \nc{\betat}{\td{\beta}}   \nc{\betah}{\hat{\beta}} 
\nc{\gammab}{\bar{\gamma}} \nc{\gammat}{\td{\gamma}} \nc{\gammah}{\hat{\gamma}} 
\nc{\deltab}{\bar{\delta}} \nc{\deltat}{\td{\delta}} \nc{\deltah}{\hat{\delta}} 
\nc{\epsilonb}{\bar{\eps}} \nc{\epsilont}{\td{\eps}} \nc{\epsilonh}{\hat{\eps}} 
\nc{\vepsb}{\bar{\veps}}   \nc{\vepst}{\td{\veps}}   \nc{\vepsh}{\hat{\veps}} 
\nc{\zetab}{\bar{\zeta}}   \nc{\zetat}{\td{\zeta}}   \nc{\zetah}{\hat{\zeta}} 
\nc{\etab}{\bar{\eta}}     \nc{\etat}{\td{\eta}}     \nc{\etah}{\hat{\eta}} 
\nc{\thetab}{\bar{\theta}} \nc{\thetat}{\td{\theta}} \nc{\thetah}{\hat{\theta}} 
\nc{\vthetab}{\bar{\vtht}} \nc{\vthetat}{\td{\vtht}} \nc{\vthetah}{\hat{\vtht}} 
\nc{\lambdab}{\bar{\lambda}} \nc{\lambdat}{\td{\lambda}} \nc{\lambdah}{\hat{\lambda}} 
\nc{\iotab}{\bar{\iota}}   \nc{\iotat}{\td{\iota}}   \nc{\iotah}{\hat{\iota}} 
\nc{\kappab}{\bar{\kappa}} \nc{\kappat}{\td{\kappa}} \nc{\kappah}{\hat{\kappa}} 
\nc{\lmdb}{\bar{\lmd}}     \nc{\lmdt}{\td{\lmd}}     \nc{\lmdh}{\hat{\lmd}} 
\nc{\mub}{\bar{\mu}}       \nc{\mut}{\td{\mu}}       \nc{\muh}{\hat{\mu}} 
\nc{\nub}{\bar{\nu}}       \nc{\nut}{\td{\nu}}       \nc{\nuh}{\hat{\nu}} 
\nc{\xib}{\bar{\xi}}       \nc{\xit}{\td{\xi}}       \nc{\xih}{\hat{\xi}} 
\nc{\pib}{\bar{\pi}}       \nc{\pit}{\td{\pi}}       \nc{\pih}{\hat{\pi}} 
\nc{\vpib}{\bar{\vpi}}     \nc{\vpit}{\td{\vpi}}     \nc{\vpih}{\hat{\vpi}} 
\nc{\rhob}{\bar{\rho}}     \nc{\rhot}{\td{\rho}}     \nc{\rhoh}{\hat{\rho}} 
\nc{\vrhob}{\bar{\vrho}}   \nc{\vrhot}{\td{\vrho}}   \nc{\vrhoh}{\hat{\vrho}} 
\nc{\sigmab}{\bar{\sigma}} \nc{\sigmat}{\td{\sigma}} \nc{\sigmah}{\hat{\sigma}} 
\nc{\vsigmab}{\bar{\vsgm}} \nc{\vsigmat}{\td{\vsgm}} \nc{\vsigmah}{\hat{\vsgm}} 
\nc{\taub}{\bar{\tau}}     \nc{\taut}{\td{\tau}}     \nc{\tauh}{\hat{\tau}} 
\nc{\upsilonb}{\bar{\ups}} \nc{\upsilont}{\td{\ups}} \nc{\upsilonh}{\hat{\ups}} 
\nc{\phib}{\bar{\phi}}     \nc{\phit}{\td{\phi}}     \nc{\phih}{\hat{\phi}} 
\nc{\varphib}{\bar{\vphi}}   \nc{\varphit}{\td{\vphi}}   \nc{\varphih}{\hat{\vphi}} 
\nc{\chib}{\bar{\chi}}     \nc{\chit}{\td{\chi}}     \nc{\chih}{\hat{\chi}} 
\nc{\psib}{\bar{\psi}}     \nc{\psit}{\td{\psi}}     \nc{\psih}{\hat{\psi}} 
\nc{\omegab}{\bar{\omega}} \nc{\omegat}{\td{\omega}} \nc{\omegah}{\hat{\omega}} 

\nc{\Gammab}{\wbar{\Gamma}}     \nc{\Gammat}{\wtd{\Gamma}}     \nc{\Gammah}{\wht{\Gamma}}
\nc{\Deltab}{\wbar{\Delta}}     \nc{\Deltat}{\wtd{\Delta}}     \nc{\Deltah}{\wht{\Delta}}
\nc{\Thetab}{\wbar{\Theta}}     \nc{\Thetat}{\wtd{\Theta}}     \nc{\Thetah}{\wht{\Theta}}
\nc{\Lambdab}{\wbar{\Lambda}}   \nc{\Lambdat}{\wtd{\Lambda}}   \nc{\Lambdah}{\wht{\Lambda}}
\nc{\Xib}{\wbar{\Xi}}           \nc{\Xit}{\wtd{\Xi}}           \nc{\Xih}{\wht{\Xi}}
\nc{\Pib}{\wbar{\Pi}}           \nc{\Pit}{\wtd{\Pi}}           \nc{\Pih}{\wht{\Pi}}
\nc{\Sigmab}{\wbar{\Sigma}}     \nc{\Sigmat}{\wtd{\Sigma}}     \nc{\Sigmah}{\wht{\Sigma}}
\nc{\Upsilonb}{\wbar{\Upsilon}} \nc{\Upsilont}{\wtd{\Upsilon}} \nc{\Upsilonh}{\wht{\Upsilon}}
\nc{\Phib}{\wbar{\Phi}}         \nc{\Phit}{\wtd{\Phi}}         \nc{\Phih}{\wht{\Phi}}
\nc{\Psib}{\wbar{\Psi}}         \nc{\Psit}{\wtd{\Psi}}         \nc{\Psih}{\wht{\Psi}}
\nc{\Omegab}{\wbar{\Omega}}     \nc{\Omegat}{\wtd{\Omega}}     \nc{\Omegah}{\wht{\Omega}}

\makeatletter
\def\wbar{\accentset{{\cc@style\underline{\mskip12mu}}}}
\def\wbarl{\accentset{{\cc@style\mskip-2mu\underline{\mskip12mu}}}}
\let\wb@r\wbar
\let\wb@rl\wbarl
\renewcommand{\wbar}[1]{\wb@r{#1}}
\renewcommand{\wbarl}[1]{\wb@rl{#1}}
\renewcommand{\Tb}{\wbarl{T}}
\renewcommand{\Qb}{\wbarl{Q}}
\makeatother

\renewcommand{\psit}{\tilde\psi}
\renewcommand{\psib}{\bar\psi}

\title{\bf Chiral algebras of $\boldsymbol{(0,2)}$ models}

\author{\\ \bf \normalsize Junya Yagi \\[5ex]
\normalsize Department of Physics, National University of Singapore \\
\normalsize  2 Science Drive 3, Singapore 117551 \\
\normalsize \tt yagi@nus.edu.sg}
\date{}

\renewcommand{\fh}{\widehat{f}}
\renewcommand{\phih}{\widehat{\phi}}
\newcommand{\afg}{\widehat{\mathfrak{g}}}

\renewcommand{\vee}{*}
\renewcommand{\CP}{\mathbb{P}}

\begin{document}
\maketitle

\begin{abstract}
  We explore two-dimensional sigma models with $(0,2)$ supersymmetry
  through their chiral algebras.  Perturbatively, the chiral algebras
  of $(0,2)$ models have a rich infinite-dimensional structure
  described by the cohomology of a sheaf of chiral differential
  operators.  Nonperturbatively, instantons can deform this structure
  drastically.  We show that under some conditions they even
  annihilate the whole algebra, thereby triggering the spontaneous
  breaking of supersymmetry.  For a certain class of K\"ahler
  manifolds, this suggests that there are no harmonic spinors on their
  loop spaces and gives a physical proof of the H\"ohn--Stolz
  conjecture.
\end{abstract}

\section{Introduction}

Supersymmetric sigma models in two dimensions have played central
roles in a number of important physical and mathematical developments
during the past few decades.  A key concept underlying much of the
developments is that of chiral rings of sigma models with $(2,2)$
supersymmetry \cite{Lerche:1989uy}.  These finite-dimensional
cohomology rings are basic ingredients of topological sigma models
\cite{Witten:1988xj, Witten:1991zz}, and intimately connected to,
among other things, Gromov--Witten invariants \cite{Witten:1988xj,
  MR809718}, Floer homology \cite{MR933228, MR948771, MR965228,
  MR987770, MR1001276}, and mirror symmetry \cite{Lerche:1989uy,
  Dixon:1987bg, Hori:2000kt}.

While the chiral rings of $(2,2)$ models are clearly very interesting,
we also know that some of the beautiful structures of two-dimensional
supersymmetric sigma models arise in essentially infinite-dimensional
contexts.  For example, the elliptic genera encode infinite series of
topological invariants of the target space \cite{Witten:1986bf,
  Witten:1987cg}.  It is then natural to ask whether there are
infinite-dimensional analogs of the chiral rings.  The answer is
``yes''.  They are the \emph{chiral algebras} of sigma models with
$(0,2)$ supersymmetry \cite{Vafa:1989pa, Witten:1993jg}.

The chiral algebra of a $(0,2)$ model is the cohomology of local
operators with respect to one of the supercharges, graded by the
right-moving R-charge and equipped with operator product expansion
(OPE).  As a consequence of $(0,2)$ supersymmetry, its elements vary
holomorphically on the worldsheet and form a structure analogous to
the chiral algebra of a conformal field theory (CFT), the operator
product algebra of holomorphic fields.  It is well known that one can
twist $(2,2)$ models to obtain topological field theories
characterized by their chiral rings.  For $(0,2)$ models, one can
perform a similar quasi-topological twisting that turns them into
holomorphic field theories characterized by their chiral algebras.

Classically, the chiral algebra of a twisted $(0,2)$ model is
isomorphic, as a graded vector space, to the direct sum of the
Dolbeault cohomology groups of a certain infinite series of
holomorphic vector bundles over the target space.  Quantum
mechanically, this isomorphism gets deformed by quantum corrections.
Like the chiral rings, the chiral algebra is independent of the choice
of the metric on the target space, hence can be computed in the large
volume limit where the theory is weakly coupled and the path integral
localizes to instantons.  But unlike the chiral rings, it receives
perturbative corrections as well as instanton corrections, because the
contributions from bosonic and fermionic fluctuations do not cancel
due to the lack of left-moving supersymmetry.  This complication leads
to the interesting subject of perturbative chiral algebras.

At the level of perturbation theory, the physics of a sigma model is
governed by the local geometry of the target space.  On the other
hand, we can always deform the target space metric to make it locally
flat without affecting the chiral algebra.  Combining these
observations, we reach a surprising conclusion: the perturbative
chiral algebra can be described locally by a sigma model with flat
target space, therefore reconstructed by gluing locally defined
\emph{free} theories globally over the target space.

This fact was exploited by Witten \cite{Witten:2005px} to show that in
the absence of left-moving fermions, the perturbative chiral algebra
of a twisted $(0,2)$ model can be formulated as the cohomology of a
sheaf of chiral differential operators on the target space, a notion
introduced earlier in mathematics by Malikov et
al.~\cite{Malikov:1998dw}.  In this picture, the moduli of the
perturbative chiral algebra are encoded in the different possible ways
of gluing relevant free CFTs, whereas the anomalies of the theory
manifest themselves in the obstructions to doing so consistently.
When left-moving fermions are present and take values in the tangent
bundle of the target space, it was shown by Kapustin
\cite{Kapustin:2005pt} that the perturbative chiral algebra is given
by the cohomology of of the chiral de Rham complex
\cite{Malikov:1998dw}.  The theory of perturbative chiral algebras has
been further developed along these lines by Tan \cite{Tan:2006qt,
  Tan:2006by, Tan:2006zg, Tan:2007bh}.

Nonperturbatively, instantons can change the picture radically
\cite{Witten:2005px, Tan-Yagi-1, Tan-Yagi-2, MR2415553}.  A
particularly striking example is the model with no left-moving
fermions whose target space is the flag manifold $G/B$ of a complex
simple Lie group $G$.  The perturbative chiral algebra of this model
is infinite-dimensional and has the structure of a $\afg$-module of
critical level \cite{Malikov:1998dw, MR1042449, MR2290768}.  In the
presence of instantons, however, the equation $1 = 0$ holds in the
cohomology and the chiral algebra \emph{vanishes}.

One clue to the existence of such a nonperturbative phenomenon lies in
a conjecture made independently by H\"ohn and Stolz \cite{MR1380455}
in the mid 1990s.  The H\"ohn--Stolz conjecture asserts that the
elliptic genus of a supersymmetric sigma model with no left-moving
fermions vanishes if the target space $M$ admits a Riemannian metric
of positive Ricci curvature.

Stolz gave a heuristic argument for his conjecture based on the
geometry of the loop space $\CL M$, the space of smooth maps from the
circle $S^1$ to $M$.  It goes as follows.  Let us assume that the
scalar curvature of $\CL M$ is given, at each loop $\gamma \in \CL M$,
by the integral of the Ricci curvature of $M$ along $\gamma$.  Then,
$\CL M$ has positive scalar curvature if $M$ has positive Ricci
curvature.  By analogy with the Lichnerowicz theorem, this would imply
that $\CL M$ has no harmonic spinors.  Meanwhile, supersymmetric
states of the theory may be identified with harmonic spinors on $\CL
M$.  Hence, the theory would have no supersymmetric states, and since
the elliptic genus counts the number of bosonic supersymmetric states
minus the number of fermionic ones at each energy level, it would
vanish then.

If Stolz's reasoning is correct, the positivity of the Ricci curvature
implies not only the vanishing of the elliptic genus, but also the
spontaneous breaking of supersymmetry.  Flag manifolds have positive
Ricci curvature, so supersymmetry should be broken in the models into
these spaces.  Supersymmetry breaking is indeed triggered whenever the
chiral algebra vanishes --- instantons tunnel between infinitely many
perturbative supersymmetric states and lift all of them at once.

In fact, what happens for the flag manifold model is a special case of
a more general phenomenon: the chiral algebra of a $(0,2)$ model with
no left-moving fermions vanishes nonperturbatively if the target space
is a compact K\"ahler manifold with positive first Chern class and
contains an embedded $\CP^1$ with trivial normal bundle.

This vanishing theorem --- or rather, ``theorem'' with quotation marks
--- is the main result of this paper, which we will ``prove'' by a
physical argument.  As a ``corollary'', the ``theorem'' implies that
supersymmetry is spontaneously broken.  Therefore, for this particular
class of K\"ahler manifolds, it suggests that there are no harmonic
spinors on their loop spaces and gives a physical proof of the
H\"ohn--Stolz conjecture.

The paper is organized as follows.  In Section 2, we introduce the
chiral algebras of $(0,2)$ models and discuss their general
properties.  Section 3 is devoted to the sheaf theory of perturbative
chiral algebras.  Finally, in Section 4, we establish the vanishing
``theorem'' and explain its relation to supersymmetry breaking and the
geometry of loop spaces.

\section{Chiral algebras of $\boldsymbol{(0,2)}$ models}
\label{CA}

We now begin our study of the chiral algebras of $(0,2)$ models.  The
focus of this section is on general properties of these algebras that
do not depend on specific details of the target space geometry.

\subsection{$\boldsymbol{(0,2)}$ models}
\label{CA-models}

Two-dimensional sigma models have $(0,2)$ supersymmetric extension
when the target space is strong K\"ahler with torsion (strong KT)
\cite{Witten:2005px}.  A Hermitian manifold is called strong KT if the
$(1,1)$-form $\omega$ associated to the Hermitian metric satisfies
$\del\delb\omega = 0$.  K\"ahler manifolds are strong KT since the
K\"ahler form satisfies $\del\omega = \delb\omega = 0$.  In this paper
we will only consider $(0,2)$ models with K\"ahler target spaces.  Let
us review how these models are constructed.

Let $\Sigma$ be a Riemann surface and $X$ a K\"ahler manifold of
complex dimension $d$.  The bosonic sigma model with worldsheet
$\Sigma$ and target space $X$ is a quantum field theory of maps
$\phi\colon \Sigma \to X$.  Imposing $(0,2)$ supersymmetry requires
the introduction of two right-moving fermions, $\psi_+$ and $\psib_+$.
These are worldsheet spinors with values in the holomorphic and
antiholomorphic tangent bundles of $X$:
\begin{equation}
  \psi_+  \in \Gamma(\Kb_\Sigma^{1/2} \otimes \phi^*T_X) \, , \quad
  \psib_+ \in \Gamma(\Kb_\Sigma^{1/2} \otimes \phi^*\Tb_X) \, .
\end{equation}
Here $\Kb_\Sigma^{1/2}$ is a square root of the antiholomorphic
canonical bundle of $\Sigma$.  The simplest $(0,2)$ model is
constructed with this minimally $(0,2)$ supersymmetric field content.

In the field space, $(0,2)$ supersymmetry is realized as the
transformation
\begin{equation}
  \label{susy}
  \begin{alignedat}{3}
    \delta\phi^i     &= -\epsilon_-\psi_+^i \, , &\quad
    \delta\phi^\ib   &= \epsilonb_-\psib_+^\ib \, , \\
    \delta\psi_+^i   &= i\epsilonb_-\del_\zb\phi^i \, , &
    \delta\psib_+^\ib &= -i\epsilon_-\del_\zb\phi^\ib \, ,
  \end{alignedat}
\end{equation}
where $\epsilon_-$ and $\epsilonb_-$ are sections of
$\Kb_\Sigma^{-1/2}$.  We define the right-moving supercharges $Q_+$
and $\Qb_+$ so that $-i\epsilon_- Q_+ + i\epsilonb_- \Qb_+$ acts by
this transformation.  The supercharges then satisfy
\begin{equation}
  \label{screl}
  \begin{gathered}
    \{Q_+, Q_+\} = \{\Qb_+, \Qb_+\} = 0 \, , \\
    \{Q_+,\Qb_+\} = -i\del_\zb = \frac{1}{2} (H - P) \, ,
  \end{gathered}
\end{equation}
and generate the $(0,2)$ supersymmetry algebra together with the
generators $H$, $P$ of translations, $M$ of rotations, and $F_R$ of
the right-moving $\U(1)$ R-symmetry.  Under the last symmetry,
$\psi_+$ has charge $-1$ and $\psib_+$ has charge $+1$; thus $Q_+$ has
charge $-1$ and $\Qb_+$ has charge $+1$.

To construct a $(0,2)$ supersymmetric action, we choose a K\"ahler
metric $g$ on $X$ and make the operator $g_{i\jb} \psi_+^i
\del_z\phi^\jb$ of R-charge $-1$.  Then the action
\begin{equation}
  \label{S1}
  S = \int_\Sigma d^2z
      \{\Qb_+, g_{i\jb} \psi_+^i \del_z\phi^\jb\}
\end{equation}
is invariant under the R-symmetry and, by virtue of the relation
$\Qb_+^2 = 0$, under the symmetry generated by $i\epsilonb_-\Qb_+$
provided that $\epsilonb_-$ is antiholomorphic and so commutes with
the $\del_z$ inside.  This action is also invariant under
$-i\epsilon_-\Qb_+$ for antiholomorphic $\epsilon_-$, as becomes clear
if we rewrite it as
\begin{equation}
  \label{S2}
  S = \int_\Sigma d^2z
      \{Q_+, g_{i\jb} \del_z\phi^i \psib_+^\jb\} \, .
\end{equation}
Expanding the anticommutators and using the K\"ahler condition, one
can check that the two expressions \eqref{S1} and \eqref{S2} both
coincide with
\begin{equation}
  \label{S}
  S = \int_\Sigma d^2z (
      g_{i\jb} \del_\zb\phi^i \del_z\phi^\jb
      + ig_{i\jb} \psi_+^i D_z\psib_+^\jb) \, .
\end{equation}
Here the covariant derivative $D_z$ is the $\del$ operator coupled to
the pullback of the Levi-Civita connection $\Gamma$ on $X$.
Explicitly, $D_z\psib_+^\ib = \del_z\psib_+^\ib + \del_z\phi^\jb
\Gamma_{\jb\kb}^\ib\psib_+^\kb$.

We can add a topological invariant to the action.  For a closed
two-form $B$ on $X$, the functional
\begin{equation}
  \label{SB}
  S_B = \int_\Sigma \phi^* B
\end{equation}
depends only on the cohomology class of $B$ and the homotopy class of
$\phi$.  As such, it is invariant under any continuous
transformations, especially the supersymmetry transformation and the
R-symmetry.  The topological invariant $S_B$ vanishes at the level of
perturbation theory, where one deals with homotopically trivial maps,
but affects the dynamics nonperturbatively.

We have obtained a $(0,2)$ supersymmetric action.  To complete the
construction, we need to make sure that a sensible quantum theory
based on this action exists.  It turns out that $X$ must satisfy two
topological conditions for that.  First, $X$ must be spin, or
equivalently, its first Chern class must be even:
\begin{equation}
  \label{spin}
  c_1(X) \equiv 0 \quad (\text{mod 2}) \, .
\end{equation}
As we will see, this condition ensures that the fermion parity
$(-1)^{F_R}$ is well defined.  Second, the second Chern character
$\ch_2(X) = c_1(X)^2/2 - c_2(X)$, which is also a half of the first
Pontryagin class $p_1(X)$, must be zero:
\begin{equation}
  \label{p1}
  \frac{1}{2} p_1(X) = 0 \, .
\end{equation}
This is the condition for the absence of sigma model anomaly
\cite{Moore:1984dc, Moore:1984ws}, the obstruction to finding a
well-defined path integral measure.  From the viewpoint of the loop
space $\CL X $, the first condition means that $\CL X$ is orientable
\cite{Witten:1985mj, MR0816738}, while the second condition (given the
first) is interpreted as the condition for $\CL X$ to admit spinors
\cite{Killingback:1986rd}.

There is also a geometric condition.  The renormalization group
generates a flow of the target space metric.  At the one-loop level,
the metric $g(\mu)$ renormalized at an energy scale $\mu$ obeys the
equation
\begin{equation}
  \label{RG}
  \mu \frac{d}{d\mu} g_{i\jb}(\mu)
  = \frac{1}{2\pi} R_{i\jb}\bigl(g(\mu)\bigr) \, ,
\end{equation}
where $R_{i\jb}(g)$ is the Ricci curvature of $g$
\cite{Friedan:1980jf, Friedan:1980jm, AlvarezGaume:1981hn,
  Callan:1985ia}.  From this equation, we see that $g(\mu)$ gets
larger as $\mu$ gets larger if the Ricci curvature is positive.  Since
the theory is weakly coupled when the target space has large volume,
we have asymptotic freedom in this case.  In contrast, if the Ricci
curvature is negative, the theory is strongly coupled for large $\mu$
and does not have a well-defined ultraviolet limit.  Hence, for the
microscopic theory to exist, the Ricci curvature should be
semipositive.  

When the above conditions are satisfied, the action \eqref{S} plus
\eqref{SB} defines the simplest version of $(0,2)$ models.  The
supercharges are given by
\begin{equation}
  Q_+ = \oint \! d\zb \, g_{i\jb} \psi_+^i \del_\zb\phi^\jb \, , \quad
  \Qb_+ = \oint \! d\zb \, g_{i\jb} \del_\zb\phi^i \psib_+^\jb
\end{equation}
and satisfy the reality condition $Q_+^\dagger = \Qb_+$.

If one has a holomorphic vector bundle $E$ over $X$, one can extend
the model by adding left-moving fermions with values in $E$.  This
extended model is called the heterotic model.  Since the left-movers
contribute to sigma model anomalies in the opposite way as the
right-movers do, their presence changes the anomaly cancellation
condition to
\begin{equation}
  \frac{1}{2} p_1(X) = \frac{1}{2} p_1(E) \, .
\end{equation}
This is trivially satisfied if $E = T_X$, in which case the model
actually has $(2,2)$ supersymmetry.  It is also possible to add
superpotentials \cite{Witten:1993yc}.  For brevity, we will not
consider these extensions in this paper.  We refer to Tan
\cite{Tan:2006qt} for perturbative aspects of the heterotic model.

\subsection{Chiral algebras}
\label{CA-CA}

Having formulated $(0,2)$ models, let us introduce the notion of their
chiral algebras.  Among the two supercharges we will use $\Qb_+$
exclusively, so simply write $Q = \Qb_+$.  Then $Q^\dagger = Q_+$.

Consider the action of $Q$ on local operators $\CO$ given by
supercommutator $[Q, \CO\}$.  The $Q$-action increases the R-charge of
$\CO$ by one, and squares to zero.  Therefore, given a $(0,2)$ model,
we can define the $Q$-cohomology of local operators graded by the
R-charge.

Since $\del_\zb \propto H - P$ is $Q$-exact by the $(0,2)$
supersymmetry algebra, it acts trivially in the $Q$-cohomology: if
$\CO$ is $Q$-closed, then $\del_\zb\CO$ is $Q$-exact.  Thus
$Q$-cohomology classes vary holomorphically on $\Sigma$.  Moreover,
two classes can be multiplied by $[\CO] \cdot [\CO'] = [\CO\CO']$.
From these facts, it follows that the $Q$-cohomology of local
operators inherits a holomorphic OPE structure from the underlying
theory:
\begin{equation}
  \label{OPE}
  [\CO(z)] \cdot [\CO'(w)]
  \sim \sum_k c_k(z - w) [\CO_k(w)] \, .
\end{equation}
The coefficient functions $c_k(z - w)$ are holomorphic away from the
diagonal $z = w$ where there can be poles.

The holomorphic $Q$-cohomology of local operators, equipped with this
natural OPE structure, is the chiral algebra of the $(0,2)$ model.  We
will denote the chiral algebra by $\CA$, and its R-charge $q$ subspace
by $\CA^q$.  As is clear from the construction, the chiral algebra of
a $(0,2)$ model has the structure of a chiral algebra in the sense of
CFT, except that the grading by conformal weight is missing.  We will
see later that it carries a similar (but possibly reduced to $\Z_n$)
grading after the theory is twisted.

The chiral algebra forms a closed sector of a $(0,2)$ model in the
following sense.  Consider the $n$-point function of $Q$-closed local
operators:
\begin{equation}
  \label{nPF}
  \bigvev{\CO_1(z_1,\zb_1) \dotsm \CO_n(z_n,\zb_n)} \, .
\end{equation}
If one of the operators is $Q$-exact, $\CO_i = [Q, \CO_i'\}$, then the
$n$-point function becomes $\pm\vev{[Q, \CO_1 \dotsm \CO_i' \dotsm
  \CO_n\}}$.  Computed with a $Q$-invariant action and path integral
measure, this is the integral of a ``$Q$-exact form'' on the field
space and vanishes.  The $n$-point function \eqref{nPF} thus depends
only on the $Q$-cohomology classes $[\CO_i]$.  In particular, it is a
holomorphic (more precisely, meromorphic) function of the insertion
points.

So far, we have considered the chiral algebra of a fixed $(0,2)$
model, defined by a fixed $Q$-invariant action.  Of course, different
choices of the action lead to different chiral algebras in general.
Imagine deforming the theory by perturbing the action:
\begin{equation}
  S \to S + \delta S \, .
\end{equation}
For this deformation to preserve the $Q$-invariance, $\delta S$ must
be $Q$-closed.  If, however, $\delta S$ is $Q$-exact, the chiral
algebra remains unchanged.  To see this, express the matrix elements
of $Q$ as path integrals on a cylinder of infinitesimal length, with a
contour integral of the supercurrent sandwiched between various
boundary conditions.  One can show that for a $Q$-exact perturbation,
the matrix elements of $Q$ in the deformed theory are equal to those
of $Q + [Q, \delta A]$ in the original theory for some operator
$\delta A$.  But this latter operator is related to $Q$ by the
conjugation
\begin{equation}
  Q \to e^{-\delta A} Q e^{\delta A} \, ,
\end{equation}
hence defines an isomorphic chiral algebra.  The $n$-point function
\eqref{nPF} is also unchanged because the perturbation just introduces
$Q$-exact insertions.

Looking back at the action \eqref{S1} of our model, we see that
deformations of the target space metric just give $Q$-exact
perturbations.  Therefore, the chiral algebra is independent of the
K\"ahler structure of the target space.  It does depend on the complex
structure, however.  For this enters the very definition of the
supersymmetry transformation.

\subsection{Instantons}
\label{CA-instantons}

An important property of the chiral algebra is that it receives
contributions only from instantons and small fluctuations around them.
In the case of our model, instantons obey
\begin{equation}
  \{Q, \psi_+^i\} = \del_\zb\phi^i = 0 \, ,
\end{equation}
so they are holomorphic maps from $\Sigma$ to $X$.  It is this
localization principle that makes the chiral algebra effectively
computable.

To prove the localization, we rescale the target space metric and make
it very large.  In this large volume limit, the path integral
localizes to the zeros of the bosonic action
\begin{equation}
  \int_\Sigma d^2z \, g_{i\jb} \del_\zb\phi^i \del_z\phi^\jb \, ,
\end{equation}
namely holomorphic maps, or instantons, as promised.  Incidentally,
there is another situation in which the same localization arises.
That is when the path integral computes the correlation function of
$Q$-closed operators \cite{Witten:1991zz}.  This situation is not
really relevant for us, though.  In order to determine the chiral
algebra, we have to ask, in the first place, whether a given local
operator is $Q$-closed or not.

The space $\CM$ of holomorphic maps from $\Sigma$ to $X$ is called the
instanton moduli space.  It decomposes into disconnected components
labeled by the homology class $\beta$ that the image of $\Sigma$
represents in $X$:
\begin{equation}
  \CM = \bigoplus_{\beta \in H_2(X,\Z)} \CM_\beta \, .
\end{equation}
Correspondingly, the path integral splits into distinct sectors each
of which integrates over the neighborhood of a single connected
component of $\CM$.  In sigma model perturbation theory, one expands
in the inverse volume of the target space in the zero-instanton
sector.  Instantons of $\beta = 0$ are constant maps, whose moduli
space $\CM_0 \iso X$.

When $c_1(X) \neq 0$, instantons induce two important nonperturbative
effects.  One is the violation of the R-charge conservation, which
breaks the R-symmetry down to a discrete subgroup.  The other is the
appearance of powers of a dynamical scale $\Lambda$ generated via
dimensional transmutation, which allows objects of different scaling
dimensions to show up as quantum corrections.

The anomaly in the R-symmetry is due to a nontrivial transformation of
the fermionic path integral measure under this symmetry.  The complex
conjugate of $\psib_+$ is a section of $K_\Sigma^{1/2} \otimes
\phi^*T_X$, while $\psi_+$ can be identified with a $(0,1)$-form with
values in the same bundle.  Thus, we can define $\Db = D_\zb d\zb$ and
expand these fields in the eigenmodes of $\Db^*\Db$ and $\Db\Db^*$:
\begin{equation}
  \psi_+  = \sum_s b_0^s v_{0,s} + \sum_n b^n v_n \, , \quad
  \psib_+ = \sum_r c_0^r \ub_{0,r} + \sum_n c^n \ub_n \, ,
\end{equation}
Here, $\ub_{0,r}$, $v_{0,s}$ are zero modes, $\ub_n$, $v_n$ nonzero
modes, and $b_0^s$, $c_0^r$, $b^n$, $c^n$ anticommuting coefficients.
The fermionic path integral measure is the formal product
\begin{equation}
  \label{FPM}
  \prod_{r, s, n} db_0^s \, dc_0^r \, db^n dc^n \, .
\end{equation}
The nonzero mode part of the measure is neutral under the R-symmetry
because of the pairing of nonzero modes.  The zero mode part has
R-charge equal to the number of $\psi_+$ zero modes minus the number
of $\psib_+$ zero modes, i.e., minus the index of the $\Db$
operator.  For a compact Riemann surface $\Sigma$, the index is given
by
\begin{equation}
  \label{index}
  \int_\Sigma \phi^*c_1(X)
\end{equation}
which (recalling $c_1(X) \equiv 0$ mod $2$) is equal to $2k$ for some
integer $k$.  The R-charge is violated by this amount in an instanton
background.  We see that the R-symmetry is broken to a $\Z_{2n}$
subgroup in the presence of instantons, where $n$ is the greatest
common divisor of the integers $k$.

Instantons produce powers of $\Lambda$ when $c_1(X) \neq 0$ because
the $B$ field is renormalized as
\begin{equation}
  \label{BREN}
  [B(\mu)] = [B_0] + \ln\frac{\mu}{\Lambda} c_1(X) \, ,
\end{equation}
with $[B_0]$ a fixed class in $H^2(X,\C)$.  We will derive this
formula at the end of this section.  Accordingly, instanton
corrections violating R-charge by $2k$ units are weighted by the
topological factor
\begin{equation}
  \label{e-SB}
  e^{-S_B}
  =\Bigl(\frac{\Lambda}{\mu}\Bigr)^{2k}
   \exp\Bigl(\int_\Sigma \phi^* B_0\Bigr) \, .
\end{equation}
An extra factor of $\mu^{2k}$ should come from somewhere else to
cancel the dependence on $\mu$.  Altogether, these corrections are
proportional to $\Lambda^{2k}$ and can relate objects of scaling
dimensions differing by $2k$.

For instanton corrections to be under control, we should choose $B_0$
such that the factor \eqref{e-SB} is exponentially suppressed for
nonconstant holomorphic maps.  The K\"ahler form is a good choice, for
example.

\subsection{Twisting}
\label{CA-twisting}

The action of our model is invariant under supersymmetries whose
transformation parameters are antiholomorphic sections of the bundle
$\Kb_\Sigma^{-1/2}$.  If $\Sigma$ is topologically nontrivial, this
bundle may not admit any global antiholomorphic sections.  In that
case the supersymmetries exist at best locally on $\Sigma$, hence so
does the chiral algebra.  However, what we need to define the chiral
algebra is really one of the two supersymmetries, not both.  So it
would be nice if we can somehow globalize one in return for giving up
the other.  This is achieved by twisting the theory.

Let us modify the spins of the fermionic fields so that $\psi_+$
becomes a $(0,1)$-form on $\Sigma$, while $\psib_+$ becomes a
zero-form.  To make this point clear, we rename them as
\begin{equation}
  -\psi_+^i \to \rho_\zb^i \, , \quad
  -i\psib_+^\ib \to \alpha^\ib \, .
\end{equation}
The action of $Q$ is then given by
\begin{equation}
  \label{Q}
  \begin{alignedat}{3}
    [Q,\phi^i]   &= 0 \, , &\quad
    [Q,\phi^\ib]  &= \alpha^\ib \, , \\
    \{Q,\rho_\zb^i\}   &= -\del_\zb\phi^i \, , &
    \{Q,\alpha^\ib\}  &= 0 \, .
  \end{alignedat}
\end{equation}
Thus $Q$ becomes a scalar and generates a global supersymmetry.  On
the other hand, $Q_+$ becomes a $(0,1)$-form and in general does not
exist globally.  In this way, we obtain a theory with global
supersymmetry on any Riemann surface $\Sigma$, described by the action
\begin{equation}
  S = \int_\Sigma d^2z
  \{Q, g_{i\jb} \rho_\zb^i \del_z\phi^\jb\}
  + \int_\Sigma \phi^* B \, .
\end{equation}
Discarding half the original supersymmetries also allows us to
consider complex target spaces which may or may not be K\"ahler.

Besides globalizing the supersymmetry, the twisting does one more
important thing: it makes the components $T_{z\zb}$ and $T_{\zb\zb}$
of the energy-momentum tensor $Q$-exact.  This is because the global
supersymmetry commutes with infinitesimal diffeomorphisms $\delta z =
v^z$, $\delta \zb = v^\zb$ up to a quantity involving $\del_\zb v^z$,
but not $\del_\zb v^\zb$, $\del_z v^z$, or $\del_z v^\zb$.  Hence, one
can take the variation of the action inside the $Q$-commutator when
computing these components.  Quantum mechanically the action gets
renormalized, but the renormalization can be done by $Q$-exact local
counterterms and the conclusion is unchanged.%
\footnote{This is not a precise statement.  When $\Sigma$ is curved,
  one needs to introduce a worldsheet metric to impose a meaningful
  cutoff length.  This produces an anomalous term in $T_{z\zb}$
  proportional to the Ricci curvature of $\Sigma$
  \cite{Callan:1985ia}.  Such a c-number anomaly does not affect the
  structure of the chiral algebra.}

The generators $T_{z\zb}$, $T_{\zb\zb}$ being $Q$-exact,
antiholomorphic reparametrizations act trivially in the
$Q$-cohomology.  Therefore after the twisting the chiral algebra has
no antiholomorphic degrees of freedom --- it defines a holomorphic
field theory.

A local operator $\CO$ is said to have dimension $(n,m)$ if inserted
at the origin it transforms as
\begin{equation}
  \CO(0) \to \lambda^{-n} \lambdab^{-m} \CO(0)
\end{equation}
under the rescaling $z \to \lambda z$, $\zb \to \lambdab\zb$.  An
immediate consequence of the decoupling of antiholomorphic degrees of
freedom is that the chiral algebra is supported by local operators
with $m = 0$, because those with $m \neq 0$ transform nontrivially
under antiholomorphic rescalings.  The holomorphic dimension, $n$, is
then equal to the spin $n - m$, so it is an integer and protected from
quantum corrections (assuming that they are small).  Thus, the chiral
algebra of the twisted model is graded by the dimension as well as the
R-charge.

If $c_1(X) \neq 0$, the grading by the R-charge is violated by
instantons as we discussed already.  The same is true of the grading
by dimension.  Instanton corrections are accompanied with powers of
$\Lambda^2$.  When multiplying a local operator in the untwisted
model, $\Lambda^2$ is best thought of as a section of $K_\Sigma
\otimes \Kb_\Sigma$, having dimension $(1,1)$.  In the twisted model,
it is more natural think of it as a section of $K_\Sigma$ with
dimension $(1,0)$, compensating the change in the dimension of the
fermionic path integral measure.  Instanton corrections violating
R-charge by $2k$ units are proportional to $\Lambda^{2k}$, so violate
dimension by $k$.  The grading by dimension is therefore reduced to
$\Z_n$ nonperturbatively when that by the R-charge is reduced to
$\Z_{2n}$.

With the grading by dimension at hand, we can now write the OPE of the
chiral algebra in the form
\begin{equation}
  \label{OPE-dim}
  [\CO_i(z)] \cdot [\CO_j(w)]
  \sim \sum_{n=0}^\infty \sum_k
  \frac{\Lambda^{2n} c^{(n)}_{ijk} [\CO_k(w)]}{(z - w)^{n_i + n_j - n_k - n}}
  \, ,
\end{equation}
where $[\CO_i]$ have dimension $n_i$ and $c^{(n)}_{ijk}$ are
constants.  The $Q$-cohomology group of dimension $n$ vanishes for $n
< 0$ since there are no local operators of negative dimension
classically.

The $Q$-cohomology classes of dimension zero are special: they have
regular OPEs, as one can see by setting $n_i = n_j = 0$ above and
noting $n_k \geq 0$.  Hence, they form an ordinary ring, much like the
chiral rings of $(2,2)$ models.  This ring is called the chiral ring
of the $(0,2)$ model \cite{Witten:2005px, MR2281544}.

We have seen advantages of the twisting.  It has one drawback,
however.  Twisting the fermions amounts to tensoring with
$\Kb_\Sigma^{-1/2}$ to which the $\del$ operator couples.  This
changes the anomaly cancellation condition: $\Sigma$ and $X$ must now
satisfy
\begin{equation}
  \frac{1}{2} p_1(X) = \frac{1}{2} c_1(\Sigma) c_1(X) = 0 \, .
\end{equation}
Here $c_1(\Sigma)$ and $c_1(X)$ are pulled back to $\Sigma \times X$.
Thus, the twisting introduces an additional anomaly which affects the
choice of $\Sigma$.  If $c_1(X) \neq 0$, we must choose $\Sigma$ with
$c_1(\Sigma) = 0$, i.e., $K_\Sigma$ must be trivial.  This was
actually implicit in our discussion when we interpreted $\Lambda^2$ as
a nowhere vanishing section of $K_\Sigma$.

\subsection{Classical  chiral algebra and  quantum deformations}
\label{CA-CQCA}

To determine the chiral algebra of a given model, what one can usually
do is to first identify its elements in the classical limit, then
include quantum corrections order by order.  What guarantees the
validity of this procedure is the following principle: small quantum
corrections can only annihilate, and never create, $Q$-cohomology
classes.

This statement is justified as follows.  Local operators that are not
$Q$-closed at a given order cannot become $Q$-closed by higher-order
corrections.  So the kernel of $Q$ can never become larger.  At the
same time, the image of $Q$ can never become smaller, because adding
quantum corrections order by order defines an injection from the image
of $Q$ at a given order to the image of $Q$ at the next order.  Then
the cohomology, which is the kernel modulo the image, can only become
smaller.

In this respect, the chiral algebra is similar to the space of
supersymmetric ground states (of supersymmetric quantum mechanics,
say).  There, small quantum corrections can lift supersymmetric ground
states by giving a small positive energy, but not create ones because
they cannot push positive energy states down to zero energy as long as
there is a gap in the beginning.  Likewise, small quantum corrections
can ``lift'' $Q$-cohomology classes by making them $Q$-exact or no
longer $Q$-closed, but not create ones because of a ``gap'' to the
$Q$-closedness or non-$Q$-exactness.

The analogy goes further.  Supersymmetry yields a one-to-one
correspondence between the bosonic positive energy states and the
fermionic ones, so supersymmetric ground states are always lifted in
boson-fermion pairs.  The same principle applies here: small quantum
corrections always annihilate $Q$-cohomology classes in boson-fermion
pairs.  More precisely, one can show that when a $Q$-cohomology class
$[\CO]$ is annihilated by quantum corrections at some order, there is
another class $[\CO']$ that is annihilated together at the same order
either via the relation $[Q, \CO\} \propto \CO'$ or $\CO \propto [Q,
\CO'\}$.%
\footnote{To prove this, let $\epsilon$ be a small parameter that
  controls the strength of quantum effects, and $\CO$ a local operator
  that is nontrivial in the $Q$-cohomology to order $\epsilon^{k-1}$.
  First, suppose that $\CO$ is no longer $Q$-closed at order
  $\epsilon^k$; thus $[Q, \CO\} = \epsilon^k \CO'$ for some $\CO'$,
  and no corrections to $\CO$ can make it $Q$-closed again.  Then
  $\CO'$ is $Q$-closed, but cannot be $Q$-exact at order $\epsilon^l$
  for any $l < k$.  For if there exists $\CO''$ such that $\epsilon^l
  \CO' = [Q, \CO''\} + O(\epsilon^{l+1})$, then $[Q, \CO -
  \epsilon^{k-l} \CO''\} = O(\epsilon^{k+1})$ and $\CO -
  \epsilon^{k-l} \CO''$ is $Q$-closed to order $\epsilon^k$.  Next,
  suppose that $\CO$ becomes $Q$-exact at order $\epsilon^k$; thus
  $\epsilon^k \CO = [Q, \CO'\} + O(\epsilon^{k+1})$ for some $\CO'$.
  Then $\CO'$ is $Q$-closed to order $\epsilon^{k-1}$, but cannot be
  $Q$-exact to the same order.  For if there exist $\CO''$ and
  $\CO'''$ such that $\epsilon^l \CO' = [Q, \CO''\} + \epsilon^{l+1}
  \CO'''$ for some $l < k$, then $\epsilon^{k-1} \CO = [Q, \CO'''\} +
  O(\epsilon^k)$ and $\CO$ is already $Q$-exact at order
  $\epsilon^{k-1}$.}

Keeping the above principles in mind, let us study the chiral algebra
of the twisted model in a little more detail.

The chiral algebra is supported by local operators whose
antiholomorphic dimension $m = 0$.  Since $\rho$ and $\zb$-derivatives
of any fields have $m > 0$, these do not enter the chiral algebra.
Furthermore, we can replace $\del_z\alpha$ by other fields using the
equation of motion $D_z\alpha^\ib = 0$.  Thus, relevant local
operators are linear combinations of operators of the form
\begin{equation}
  \label{LO}
  \CO(\phi,\phib)_{j\dotsm k\dotsm \lb\dotsm  \mb\dotsm \ib_1 \dotsm \ib_q}
  \del_z\phi^j \dotsm \del_z^2\phi^k \dotsm
  \del_z\phi^\lb \dotsm \del_z^2\phi^\mb \dotsm
  \alpha^{\ib_1} \dotsm \alpha^{\ib_q} \, .
\end{equation}
Identifying $\alpha^\ib$ with the differential $d\phi^\ib$, we can
regard such operators with R-charge $q$ and dimension $n$ as
$(0,q)$-forms with values in a certain holomorphic vector bundle
$V_{X,n}$ over $X$.

For example, an $n = 0$ operator $\CO_{\ib_1\dotsm\ib_q}
\alpha^{\ib_1} \dotsm \alpha^{\ib_q}$ is a $(0,q)$-form on $X$, hence
$V_{X,0} = 1$, the trivial bundle of rank $1$.  At $n = 1$, there are
two types of operators, $\CO_{j\ib_1\dotsm\ib_q} \del_z\phi^j
\alpha^{\ib_1} \dotsm \alpha^{\ib_q}$ and $\CO^j{}_{\ib_1\dotsm\ib_q}
g_{j\kb} \del_z\phi^\kb \alpha^{\ib_1} \dotsm \alpha^{\ib_q}$.  These
are $(0,q)$-forms with values in $T_X^\vee$ and $T_X$, thus $V_{X,1} =
T_X \oplus T_X^\vee$.  At $n = 2$, we have five types, giving $V_{X,2}
= T_X \oplus T_X^\vee \oplus S^2T_X \oplus (T_X \otimes T_X^\vee)
\oplus S^2T_X^\vee$.  Here $S^k$ is the symmetric $k$th power.  In
general, $V_{X,n}$ is given by the series
\begin{equation}
  \label{Vn}
  \sum_{n = 0}^\infty q^n V_{X,n}
  = \bigotimes_{k = 0}^\infty S_{q^k}(T_X)
  \bigotimes_{l = 0}^\infty S_{q^l}(T_X^\vee) \, ,
\end{equation}
where $S_t(V) = 1 + tS(V) + t^2 S^2(V) + \dotsb$.  We will write
\begin{equation}
  V_X = \bigoplus_{n = 0}^\infty V_{X,n}.
\end{equation}

At the classical level, we can easily find the action of $Q$ on these
operators.  As an example, take $\CO_{j\ib_1\dotsm\ib_q} \del_z\phi^j
\alpha^{\ib_1} \dotsm \alpha^{\ib_q}$.  Acting on it with $Q$ gives
$\alpha^\ib \del_\ib \CO_{j\ib_1\dotsm\ib_q} \del_z\phi^j
\alpha^{\ib_1} \dotsm \alpha^{\ib_q}$.  On an $n = 1$ operator of the
other type, $\CO^j{}_{\ib_1\dotsm\ib_q} g_{j\kb} \del_z\phi^\kb
\alpha^{\ib_1} \dotsm \alpha^{\ib_q}$, acting with $Q$ gives
$\alpha^\ib \del_\ib\CO^j{}_{\ib_1\dotsm\ib_q} g_{j\kb} \del_z\phi^\kb
\alpha^{\ib_1} \dotsm \alpha^{\ib_q}$ plus $\CO^j{}_{\ib_1\dotsm\ib_q}
g_{j\kb} D_z\alpha^\kb \alpha^{\ib_1} \dotsm \alpha^{\ib_q}$, but the
latter vanishes by the equation of motion.  From these examples, we
see that $Q$ acts as the $\delb$ operator.%
\footnote{General local operators of dimension greater than one are
  constructed using covariant derivatives.  On such operators, $Q$
  acts by $\delb$ plus terms involving the curvature of the target
  space.  Classically, these additional terms which vanish for a flat
  metric do not change the $Q$-cohomology.  This point will become
  clear when we develop a sheaf theory approach to the perturbative
  chiral algebra in Section~\ref{PCA}.}
Classically, the $q$th $Q$-cohomology group of dimension $n$ is
therefore isomorphic to the $q$th Dolbeault cohomology group
$H_\delb^q(X, V_{X,n})$, and
\begin{equation}
  \label{Acl}
  \CA \iso \bigoplus_{q = 0}^d \bigoplus_{n = 0}^\infty H_\delb^q(X, V_{X,n})
\end{equation}
as graded vector spaces.  It is clear from this formula that the
chiral algebra is generally infinite-dimensional.

Quantum mechanically, the action of $Q$ is deformed by quantum
corrections.  Because of this deformation, some of the classical
$Q$-cohomology classes can disappear.

A notable example is the disappearance of the energy-momentum tensor.
Classically, our model is conformally invariant and $T_{zz}$ is
$Q$-closed on-shell.  We expect that it is no longer $Q$-closed
perturbatively if the Ricci curvature is nonzero, since conformal
invariance is broken at one loop in that case.  To determine
$[Q,T_{zz}]$, we note that $Q$ commutes with $\del_z \propto H + P$.
Thus
\begin{equation}
  \Bigl[Q, \oint \! dz \, T_{zz} - \oint \! d\zb \, T_{z\zb}\Bigr]
  = \oint \! dz [Q, T_{zz}]
  = 0 \, ,
\end{equation}
where we have used the fact that $T_{z\zb}$ is $Q$-exact.  This
suggests
\begin{equation}
  [Q, T_{zz}] = \del_z\theta
\end{equation}
for some $\theta$.  Presumably, $\theta$ is a $Q$-closed local
operator of R-charge one and dimension one, constructed from the
target space metric.  To leading order, $Q = \delb$ and for a generic
choice of the metric there is only one possibility:
\begin{equation}
  \theta \propto R_{i\jb} \del_z\phi^i\alpha^\jb \, .
\end{equation}
So, as expected, $T_{zz}$ would cease to be $Q$-closed unless the
Ricci curvature vanishes.  In Section \ref{PCA-c1}, we will compute
$[Q, T_{zz}]$ explicitly in perturbation theory and find that it is
indeed proportional to $\del_z(R_{i\jb} \del_z\phi^i\alpha^\jb)$ up to
higher-order corrections.

If $c_1(X) = 0$, then $\theta$ is $Q$-exact and we can add corrections
to $T_{zz}$ that make it $Q$-closed again.  But if $c_1(X) \neq 0$,
there is no way to do this, so $[T_{zz}]$ is annihilated together with
$[\del_z\theta]$ by perturbative corrections.  The chiral algebra is a
little exotic in this case: it is analogous to the chiral algebras of
CFTs, but lacks an energy-momentum tensor and hence invariance under
holomorphic reparametrizations.

Except for the possible lack of energy-momentum tensor, perturbatively
the structure of the chiral algebra is not very different from its
classical description.  The basic fact about perturbative corrections
is that they are local on the target space, because one only considers
fluctuations around constant maps in perturbation theory.  Moreover,
the gradings by the R-charge and dimension are not violated
perturbatively.  Thanks to these properties, at the perturbative level
the action of $Q$ still defines differential complexes
\begin{equation}
  \dotsb \longto V_{X,n} \otimes \wedge^q \Tb_X^\vee
         \stackrel{Q}{\longto} V_{X,n} \otimes \wedge^{q+1} \Tb_X^\vee
         \longto \dotsb \, ,
\end{equation}
and the chiral algebra is given by the direct sum of their cohomology
groups.  How to compute these cohomology groups using a sheaf of free
CFTs on $X$ is the subject of the next section.

Beyond perturbation theory, the physics is no longer local on the
target space.  Rather, it is local on the instanton moduli space $\CM$
since quantum fluctuations localize to instantons.  So presumably the
chiral algebra can be formulated nonperturbatively as a cohomology
theory on $\CM$, but exactly how this should be done is not clear.  At
any rate, in principle one can always compute the instanton
corrections to the action of $Q$ by path integral and determine the
exact chiral algebra.  Instanton effects often lead to surprising
results.  Later we will see examples where the whole chiral algebra is
annihilated by instanton corrections.

Let us tie up the loose ends from Section \ref{CA-instantons} by
explaining why the $B$ field should be renormalized as asserted there.
Consider the case where the twisted and untwisted models are
isomorphic.  Suppose that instantons annihilate perturbative
$Q$-cohomology classes $[\CO]$ and $[\CO']$ through a relation of the
form $[Q,\CO\} \sim \CO'$.  If this relation violates R-charge by $2k$
units, the left-hand side contains $2k$ more $\alpha$ fields than the
right-hand side does, while the both sides have antiholomorphic
dimension equal to zero.  Viewed in the untwisted model, this means
that these instantons relate operators whose antiholomorphic
dimensions differ by $k$.  Since the dynamical scale $\Lambda$ is the
only dimensional parameter available and has antiholomorphic dimension
$1/2$, to match the scaling dimensions a factor of $\Lambda^{2k}$ must
appear in the right-hand side.  For that, the $B$ field must obey the
renormalization group equation \eqref{BREN}.

\section{Sheaf theory of perturbative chiral algebras}
\label{PCA}

As we have explained in the last section, the chiral algebra can be
understood as a quantum deformation of the Dolbeault cohomology of an
infinite-dimensional holomorphic vector bundle over the target space.
At the perturbative level, there is an alternative formulation of the
chiral algebra, which involves a sheaf of free CFTs.  The goal of this
section is to develop the sheaf theory of perturbative chiral
algebras.

\subsection{Perturbative chiral algebra from free CFTs}

The chiral algebra of the twisted model is classically the Dolbeault
cohomology of the holomorphic vector bundle $V_X$.  By the \v
Cech--Dolbeault isomorphism, this cohomology is isomorphic to the
cohomology of the sheaf of holomorphic sections of $V_X$.
Perturbatively $Q$ gets corrected, but still acts as a differential
operator on the target space due to the locality of sigma model
perturbation theory.  In such a situation, there is an analog of the
\v Cech--Dolbeault isomorphism: the perturbative $Q$-cohomology is
isomorphic to the cohomology of the sheaf of perturbatively $Q$-closed
sections of $V_X$.  The proof is completely parallel to the classical
case.%
\footnote{The key ingredients are the $Q$-Poincar\'e lemma and the
  existence of partitions of unity on the sheaf of $(0,q)$-forms with
  values in $V_X$.  The former follows from the $\delb$-Poincar\'e
  lemma since small quantum corrections cannot create $Q$-cohomology
  classes.  As for the latter, ``quantum'' partitions of unity can
  always be constructed by renormalizing ``classical'' partitions of
  unity.}

This ``\v Cech-$Q$ isomorphism'' may seem to have little practical
value.  To compute the sheaf cohomology, first of all one needs to
know the general form of perturbatively $Q$-closed local sections of
$V_X$.  But this requires understanding beforehand how perturbative
corrections deform the classical expression $Q = \delb$ precisely,
which is generally very hard if not impossible.  However, we can
circumvent this difficulty if we adopt a different approach.

Let us recast the \v Cech-$Q$ isomorphism in a slightly more abstract
form.  Choose a good cover $\{U_\alpha\}$ on $X$; thus all nonempty
finite intersections of $U_\alpha$ are diffeomorphic to $\C^d$.  On
each $U_\alpha$, the space of perturbatively $Q$-closed sections of
$V_X$ is isomorphic to the chiral algebra of the twisted model into
$U_\alpha$.  For, the zeroth $Q$-cohomology group is isomorphic to
this space, while the higher $Q$-cohomology groups vanish as
$U_\alpha$ is topologically trivial.  Since \v Cech cohomology is
defined by taking the direct limit as the open cover becomes finer and
finer, and every open cover has a refinement by a good cover, it
follows that the perturbative chiral algebra can be computed via the
cohomology of the sheaf of chiral algebras $\CAh$ on $X$:
\begin{equation}
  \CA^q \iso H^q(X,\CAh) \, .
\end{equation}
Rewriting the \v Cech-$Q$ isomorphism this way makes it clear that
the perturbative chiral algebra can be formulated without reference to
any globally defined metric on the target space.

When one computes the cohomology of $\CAh$, say using a good cover
$\{U_\alpha\}$, one can endow $U_\alpha$ with any metric.  In
practice, we will put $g_{i\jb} = \delta_{i\jb}$ on all the
$U_\alpha$.  The point is that the twisted models with the flat target
spaces $U_\alpha$ are free theories --- their chiral algebras receive
no quantum corrections.

In the free twisted model $Q = \delb$ exactly, so sections of
$\CAh(U_\alpha)$ are represented by local operators of the form
$\CO(\phi, \del_z\phi, \dotsc, \del_z\phib, \del_z^2\phib, \dotsc)$.
Introducing bosonic fields $\beta_i$ of dimension one and $\gamma^i$
of dimension zero by
\begin{equation}
  \beta_i = 2\pi \delta_{i\jb}\del_z\phib^\jb \, , \quad
  \gamma^i = \phi^i \, ,
\end{equation}
we can conveniently write them as $\CO(\gamma, \del_z\gamma, \dotsc,
\beta, \del_z\beta, \dotsc)$.  These are observables of the free
$\beta\gamma$ system, a free CFT with action
\begin{equation}
  S = \frac{1}{2\pi} \int_\Sigma d^2z \, \beta_i \del_\zb\gamma^i
\end{equation}
whose OPEs are
\begin{equation}
  \beta_i(z) \gamma^j(w) \sim -\frac{\delta_i^j}{z - w} \, , \quad
  \beta_i(z) \beta_j(w) \sim 0 \, , \quad
  \gamma^i(z) \gamma^j(w) \sim 0 \, .
\end{equation}
Hence, $\CAh$ may be considered as a sheaf of free $\beta\gamma$
systems.  We conclude that the perturbative chiral algebra can be
reconstructed by gluing free $\beta\gamma$ systems over $X$ and
computing the \v Cech cohomology.

This construction is exact to all orders in perturbation theory since
the spaces $\CAh(U_\alpha)$ are free of quantum corrections.  Where
did the perturbative corrections go?  They are now encoded in the
transition functions $\fh_{\alpha\beta}$ for the sheaf $\CAh$.
Instead of equipping flat metrics on the $U_\alpha$, one may as well
start with a globally defined, curved metric on $X$.  One can then
flatten it over each $U_\alpha$ to obtain a free $\beta\gamma$ system.
In this process the perturbative corrections disappear locally, but
the transition functions change by $\fh_{\alpha\beta} \to
e^{-A_\alpha} \fh_{\alpha\beta} e^{A_\beta}$ for some operators
$A_\alpha$.  These operators carry the information on the perturbative
corrections.

But then, how can we find the $A_\alpha$?  Here we come back to the
original problem: to do that, we need detailed knowledge of the
perturbative corrections.  Instead, what we can do is take a
collection of locally defined free $\beta\gamma$ systems and glue them
using various choices of the transition functions.  In doing so, we
are effectively parametrizing the theory by the way this gluing is
done.  This is the strategy we will adopt.

\subsection{Gluing $\boldsymbol{\beta\gamma}$ systems}

We are thus led to the problem of listing the possible sets of
transition functions for $\CAh$.  The first step is to classify the
automorphisms of the free $\beta\gamma$ system.  Automorphisms are
generated by currents of dimension one.  There are two types of such
currents.

Let $V$ be a holomorphic vector field on $\C^d$.  Then $J_V =
-V^i(\gamma) \beta_i$ is a good current.  (Here and below, normal
ordering is implicit in expressions containing both $\beta_i$ and
$\gamma^i$.)  From the OPEs
\begin{equation}
  J_V(z) \gamma^i(w) \sim \frac{V^i(w)}{z - w} \, , \quad
  J_V(z) \beta_i(w) \sim -\frac{\del_iV^j \beta_j(w)}{z - w} \, ,
\end{equation}
we see that $J_V$ generates the infinitesimal diffeomorphism
$\delta\gamma = V$ and $\beta$ transforms as a $(1,0)$-form under this
symmetry.  We denote the corresponding conserved charge by $K_V$.

With a holomorphic one-form $B$ on $\C^d$, we can also make $J_B =
B_i(\gamma) \del_z\gamma^i$.  This has the OPEs
\begin{equation}
  J_B(z) \gamma^i(w) \sim 0 \, , \quad
  J_B(z) \beta_i(w)
  \sim -\frac{B_i(w)}{(z - w)^2}
       +\frac{C_{ij} \del_z\gamma^j(w)}{z - w}
\end{equation}
with $C = \del B$, so generates the transformation $\delta\beta =
-i_{\del_z\gamma} C$.  For any closed holomorphic two-form $C$, there
is a holomorphic one-form $B$ such that $C = \del B$.  Moreover,
$\delta\beta = 0$ if and only if $C = 0$.  The automorphisms of this
type are therefore labeled by the closed holomorphic two-forms $C$.
We denote their charges by $K_C$.

One can readily work out the commutators between the conserved
charges.  Computing the relevant OPEs, one finds
\begin{equation}
  \begin{split}
    [K_V, K_{V'}] &= K_{[V,V']} + K_{C(V,V')} \, , \\
    [K_V, K_C] &= K_{\CL_V C} \, , \\
    [K_C, K_{C'}] &= 0 \, .
  \end{split}
\end{equation}
Here, $C(V,V') = \del_i\del_k V^l \del_j\del_l V'^k d\phi^i \wedge
d\phi^j$.  The last two of these relations show that the $K_C$
generate an abelian subalgebra on which holomorphic reparametrizations
act naturally.

Now, suppose that the complex manifold $X$ is built up by gluing open
patches $U_\alpha \iso \C^d$ using transition functions
$f_{\alpha\beta}$.  Transition functions for $\CAh$ are constructed by
lifting $f_{\alpha\beta}$ to automorphisms $\fh_{\alpha\beta}$ of the
$\beta\gamma$ system compatible with the cocycle condition.  So we
choose $\fh_{\alpha\beta}$ such that they act on $\gamma$ by
$f_{\alpha\beta}$ and satisfy $\fh_{\alpha\beta} \fh_{\beta\gamma}
\fh_{\gamma\alpha} = 1$ on $U_\alpha \cap U_\beta \cap U_\gamma$.

Naively, one might think that the cocycle condition is satisfied if
one picks a cocycle $\{C_{\alpha\beta}\}$ of closed holomorphic
two-forms and let $\fh_{\alpha\beta}$ act on $\beta$ by the pullback
$f_{\alpha\beta}^*$ followed by $\exp(K_{C_{\alpha\beta}})$.  The
situation is actually more complicated.  The commutator between two
$K_V$s differs from the expected form by a $K_C$ term, implying
\begin{equation}
  \fh_{\alpha\beta} \fh_{\beta\gamma} \fh_{\gamma\alpha}
  = \exp(K_{C_{\alpha\beta\gamma}})
\end{equation}
for some $C_{\alpha\beta\gamma}$.  We need to adjust
$\fh_{\alpha\beta}$ so that $C_{\alpha\beta\gamma}$ disappear.  The
freedom at our disposal is to transform $\fh_{\alpha\beta} \to
\exp(K_{C'_{\alpha\beta}}) \fh_{\alpha\beta}$, which shifts
\begin{equation}
  \label{C2C+D}
  C_{\alpha\beta\gamma}
  \to C_{\alpha\beta\gamma} + (\delta C')_{\alpha\beta\gamma} \, .
\end{equation}
Unless $C_{\alpha\beta\gamma}$ can be canceled by an appropriate
choice of $C'_{\alpha\beta}$, the gluing cannot be carried out
consistently.  In other words, there may be an obstruction to the
existence of a sheaf of $\beta\gamma$ systems.

The obstruction has a natural physical interpretation.  It is not hard
to show that $C_{\alpha\beta\gamma}$ are totally antisymmetric in
$\alpha$, $\beta$, $\gamma$ and obey $(\delta
C)_{\alpha\beta\gamma\delta} = 0$, hence define a cocycle.  The
freedom \eqref{C2C+D} then means that the obstruction is encoded in
the cohomology class
\begin{equation}
  [C_{\alpha\beta\gamma}] \in H^2(X, \Omega^{2,\mathrm{cl}}_X) \, ,
\end{equation}
where $\Omega^{2,\mathrm{cl}}_X$ is the sheaf of closed holomorphic
two-forms on $X$.  It can be shown \cite{MR1748287} that this class is
mapped to $p_1(X) \in H^4(X,\R)$ under the \v Cech--Dolbeault
isomorphism.  Thus, the obstruction vanishes if and only if $p_1(X) =
0$.  This is the condition for the perturbative cancellation of sigma
model anomaly.  Since the obstruction arises in lifting the
diffeomorphisms used to construct the underlying manifold $X$, in this
way we see that the sigma model anomaly causes a gravitational anomaly
on the target space.  The cancellation of the anomaly by adjusting
$\fh_{\alpha\beta}$ is a kind of the Green--Schwarz mechanism.

Given $f_{\alpha\beta}$, the choice of $\fh_{\alpha\beta}$ is not
unique.  Suppose that we choose different transition functions,
$\fh'_{\alpha\beta}$.  Then the two sets of transition functions are
related by $\fh'_{\alpha\beta} = \exp(K_{C'_{\alpha\beta}})
\fh_{\alpha\beta}$ for some cocycle $\{C'_{\alpha\beta}\}$.  If this
cocycle is exact, $C'_{\alpha\beta} = (\delta C'')_{\alpha\beta}$, then
we have $\fh'_{\alpha\beta} = \exp(-K_{C''_\alpha}) \fh_{\alpha\beta}
\exp(K_{C''_\beta})$ and these define the same gluing.  Hence,
inequivalent choices of the transition functions are parametrized by
\begin{equation}
  H^1(X, \Omega^{2,\mathrm{cl}}_X) \, .
\end{equation}

The origin of this moduli space can be understood as follows.  For any
closed form $\CH$ of type $(3,0) \oplus (2,1)$, locally we can find a
$(2,0)$-form $T$ such that $\CH = dT$.%
\footnote{By the Poincar\'e lemma, locally $\CH = d(U + V)$ for some
  $(2,0)$-form $U$ and $\delb$-closed $(1,1)$-form $V$.  By the
  $\delb$-Poincar\'e lemma, locally $V = \delb W$ for some
  $(1,0)$-form $W$.  Then $T = U + V - dW$.}
Thus, we have the short exact sequence
\begin{equation}
  0 \longto \Omega^{2,\mathrm{cl}}_X
    \longto \CA^{2,0}_X
    \stackrel{d}{\longto} \CZ_X^{3,0} \oplus \CZ_X^{2,1}
    \longto 0 \, ,
\end{equation}
where $\CA^{p,q}_X$ and $\CZ_X^{p,q}$ are the sheaves of $(p,q)$-forms
and closed $(p,q)$-forms on $X$, respectively.  Since $H^p(X,
\CA^{2,0}_X) = 0$ for $p > 0$, the long exact sequence of cohomology
implies
\begin{equation}
  H^1(X, \Omega^{2,\mathrm{cl}}_X)
  \iso H^0(X, \CZ_X^{3,0} \oplus \CZ_X^{2,1})/dH^0(X, \CA^{2,0}_X) \, .
\end{equation}
This is the space of closed forms $\CH$ of type $(3,0) \oplus (2,1)$
modulo those that can be written as $\CH = dT'$ with a globally
defined $(2,0)$-form $T'$.  It actually parametrizes the conformally
invariant $Q$-closed term
\begin{equation}
  \int_\Sigma d^2z \{Q, T_{ij} \rho_\zb^i \del_z\phi^j\}
  = \int_\Sigma d^2z \,  \alpha^\kb \CH_{\kb ij} \rho_\zb^i \del_z\phi^j
    + i \! \int_\Sigma \phi^* T \, ,
\end{equation}
which depends on $T$ only through $\CH$ perturbatively.  We can add
this term to our action.  Then the chiral algebra is deformed unless
$T$ is globally defined.  Still, we can always set $T$ to zero locally
by subtracting a globally defined $T'$.  Combined with flattening the
metric, this turns the action locally into the free theory form.
Therefore, the effect of this term can be treated --- and is
necessarily included --- in the present framework.

Up to this point, we have implicitly worked in some coordinate
neighborhood of $\Sigma$.  The sheaf $\CAh$ of chiral algebras in this
case (or when $\Sigma \iso \C$) is known as a sheaf of chiral
differential operators \cite{Malikov:1998dw, MR1748287}.  In order to
reconstruct the chiral algebra globally on $\Sigma$, one has to glue
sheaves of chiral differential operators patch by patch over $\Sigma$.
This amounts to gluing free $\beta\gamma$ systems over $\Sigma \times
X$.  The obstruction thus takes values in
\begin{equation}
  H^2(\Sigma \times X, \Omega^{2,\mathrm{cl}}_{\Sigma \times X}).
\end{equation}
A part of it depends on both $\Sigma$ and $X$, and corresponds to the
$c_1(\Sigma) c_1(X)/2$ anomaly.  Likewise, the moduli are parametrized
by
\begin{equation}
  H^1(\Sigma \times X, \Omega^{2,\mathrm{cl}}_{\Sigma \times X}) \, .
\end{equation}

\subsection{Conformal anomaly}
\label{PCA-c1}

Previously, we claimed that the chiral algebra lacks invariance under
holomorphic reparametrizations when $c_1(X) \neq 0$, arguing that in
that case perturbative corrections would annihilate the classical
$Q$-cohomology class $[T_{zz}]$ together with another class
$[\del_z\theta]$ via the relation $[Q, T_{zz}] = \del_z\theta$.
However, we could not really exclude the possibility that $\theta$
turns out to be zero and $T_{zz}$ remains $Q$-closed.  Let us check
that this does not happen using the tool developed in this section.

Let $\{U_\alpha\}$ be a good cover of $X$.  On each $U_\alpha$, we put
a free $\beta\gamma$ system with energy-momentum tensor $T_\alpha =
-\beta_{\alpha i} \del_z\gamma_\alpha^i$.  The OPE
\begin{equation}
  J_V(z) T_\alpha(w)
  \sim -\frac{\del_i V^i(w)}{(z - w)^3}
       - \frac{(\del_z\del_i V^i + V^i\beta_i)(w)}{(z - w)^2}
       - \frac{1}{2} \frac{\del_z^2(\del_i V^i)(w)}{z - w}
\end{equation}
shows that $T_\alpha$ transform as $\delta T_\alpha = -\del_z^2\del_i
V^i/2$ under infinitesimal diffeomorphisms $\delta\gamma = V$.  The
finite form of this transformation is \cite{MR1748287}
\begin{equation}
  T_\beta - T_\alpha
  = -\frac{1}{2} \del_z^2 \log \det
    \frac{\del\gamma_\beta}{\del\gamma_\alpha} \, .
\end{equation}
Here $\del\gamma_\beta/\del\gamma_\alpha$ is the Jacobian matrix.  We
define a cocycle $\{\theta_{\alpha\beta}\}$ by
\begin{equation}
  \theta_{\alpha\beta}
  = -\frac{1}{2} \del_z \log \det
     \frac{\del\gamma_\beta}{\del\gamma_\alpha}
\end{equation}
so that it satisfies $T_\beta - T_\alpha =
\del_z\theta_{\alpha\beta}$.  Via the \v Cech-$Q$ isomorphism, this
equation translates to $[Q, T_{zz}] = \del_z\theta$ for some local
operator $\theta$.

The explicit form of $\theta$ can be found as follows.  Write
$\theta_{\alpha\beta} = W_\beta - W_\alpha$, where $W_\alpha$ is the
quantity $W = \del_z\phi^i \del_i \log \det g/2$ evaluated in the
coordinate patch $U_\alpha$.  Since $W_\alpha$ transform by
holomorphic transition functions, $\delb W$ is globally defined.  This
gives $\theta$ as a global section of the sheaf of free $\beta\gamma$
systems (on which $Q$ acts by $\delb$).  Noting $R_{i\jb} =
-\del_i\delb_\jb \log \det g$, we find
\begin{equation}
  \label{theta}
  \theta = -\frac{1}{2} R_{i\jb} \del_z\phi^i \alpha^\jb \, .
\end{equation}
As a globally defined local operator of the original theory, this
formula gets higher-order corrections since $Q = \delb$ only to
leading order.  The form of $\theta$ is in accord with the previous
discussion.

We introduced the perturbative $Q$-cohomology class $[\theta]$ through
the action of $Q$ on $T_{zz}$.  We can also construct it as follows,
mimicking the definition of the first Chern class.  Let
$f_{\alpha\beta}$ be transition functions of $K_X^\vee$.  The cocycle
condition $f_{\alpha\beta} f_{\beta\gamma} f_{\gamma\alpha} = 1$
implies that $(\delta \log f)_{\alpha\beta\gamma}$ are integer
multiples of $2\pi i$.  Thus, by applying $\delta\log/2\pi i$ on
$f_{\alpha\beta}$, we obtain an element of $H^2(X,\Z)$.  It is
$c_1(X)$.  To obtain an element of the chiral algebra, we apply
$\del_z\log/2$ instead.  This gives $\del_z\log f_{\alpha\beta}/2 =
\theta_{\alpha\beta}$, which represents $[\theta] \in H^1(X,\CAh)$.

\subsection{$\boldsymbol{\CP^1}$ model}
\label{PCA-CP1}

To conclude the discussion of the sheaf theory approach, let us
compute the perturbative chiral algebra of the twisted model with
target space $X = \CP^1$ for the first few dimensions.  We will work
locally on $\Sigma$.

The target space $\CP^1 \iso \C \cup \{\infty\}$ is covered by two
patches, $U = \CP^1 \setminus \{\infty\}$ with coordinate $\gamma$ and
$U' = \CP^1 \setminus \{0\}$ with coordinate $\gamma'$, related to
each other by
\begin{equation}
  \label{gamma'}
  \gamma' = \frac{1}{\gamma} \, .
\end{equation}
Classically $\beta$ transforms as $\beta' = -\gamma^2 \beta$, but this
formula gets corrected quantum mechanically.  The quantum
transformation law turns out to be
\begin{equation}
  \label{beta'}
  \beta' = -\gamma^2 \beta + 2\del_z\gamma \, .
\end{equation}
The additional term is needed to keep the $\beta\beta$ OPE regular
under this transformation.  Since the moduli space $H^1(\CP^1,
\Omega^{2,\mathrm{cl}}_{\CP^1}) = 0$, this is essentially the only way
to glue the two free $\beta\gamma$ systems.

We first look at the zeroth $Q$-cohomology group $\CA^0$.  The
elements of $\CA^0$ are represented by global sections of the sheaf
$\CAh$ of chiral algebras on $\CP^1$.

At dimension zero, relevant local operators are holomorphic functions.
Since a holomorphic function on a compact complex manifold must be
constant, the dimension zero subspace of $\CA^0$ is one-dimensional
and generated by the cohomology class $[1]$, represented by the
identity operator $1$.

At dimension one, the possible local operators are those of the form
$B(\gamma)\del_z\gamma$ and $V(\gamma)\beta$, where $B$ is a
holomorphic one-form and $V$ is a holomorphic vector field.  There are
no global holomorphic one-forms on $\CP^1$.  For holomorphic vector
fields, we have three independent ones, $\del$, $-\gamma\del$, and
$-\gamma^2\del$.  Hence, at the classical level we have three
cohomology classes, represented by $\beta$, $-\gamma\beta$, and
$-\gamma^2\beta$.  These survive to the perturbative chiral algebra.
Their quantum counterparts are
\begin{equation}
  \begin{split}
    J_- &= \beta = -\gamma'^2\beta' + 2\del_z\gamma' \, , \\
    J_3 &= -\gamma\beta = \gamma'\beta' \, , \\
    J_+ &= -\gamma^2\beta + 2\del_z\gamma = \beta' \, ,
  \end{split}
\end{equation}
generating the affine Lie algebra $\widehat{\mathfrak{sl}}_2$ at the
critical level $-2$:
\begin{equation}
  \begin{split}
    J_3(z) J_3(w) &\sim -\frac{1}{(z - w)^2} \, , \\
    J_3(z) J_\pm(w) &\sim \pm\frac{J_\pm(w)}{(z - w)^2} \, , \\
    J_+(z) J_-(w) &\sim -\frac{2}{(z - w)^2} + \frac{2J_3(w)}{z - w} \, .
  \end{split}
\end{equation}
The existence of these currents in the perturbative chiral algebra is
a reflection of the fact that $\CP^1$ admits an $\SL_2$-action.

We now turn to the first $Q$-cohomology group $\CA^1$.  The elements
of $\CA^1$ are represented by sections of $\CAh(U \cap U')$ that
cannot be written as the difference of a section of $\CAh(U)$ and a
section of $\CAh(U')$.  Extended over the whole target space $\CP^1$,
such sections necessarily have poles at both $0$ and $\infty$.

Meromorphic functions with poles at $0$ and $\infty$ can always be
split into a part regular at $0$ and a part regular at $\infty$.  Thus
the dimension zero subspace of $\CA^1$ is zero.

At dimension one, we can try operators of the form $\beta/\gamma^n$,
which have a pole at $0$.  However, they are all regular at $\infty$.
The other possibilities are $\del_z\gamma/\gamma^n$.  Requiring they
have a pole at $\infty$, we find that only $\del_z\gamma/\gamma$ can
represent a nontrivial cohomology class.  Indeed, it does.  This
cohomology class is $[\theta]$ since $\del_z\gamma/\gamma =
-\del_z\log(\del\gamma'/\del\gamma)/2$.

At dimension two, sections with poles at both $0$ and $\infty$ are
linear combinations of $\del_z^2\gamma/\gamma$,
$\del_z^2\gamma/\gamma^2$, $(\del_z\gamma)^2/\gamma$,
$(\del_z\gamma)^2/\gamma^2$, $(\del_z\gamma)^2/\gamma^3$, and
$\beta\del_z\gamma/\gamma$.  Among these, the combinations
$\del_z^2\gamma/\gamma^2 - 2(\del_z\gamma)^2/\gamma^3$ and
$\beta\del_z\gamma/\gamma + (\del_z\gamma)^2/\gamma^3$ are regular at
$\infty$, so vanish in the cohomology.  (In verifying this assertion,
one should keep in mind that the latter operator is normal ordered.)
Moreover, $\del_z(\del_z\gamma/\gamma)$ also vanishes due to the
perturbative relation $[Q, T_{zz}] = \del_z\theta$.  Thus the
dimension two subspace of $\CA^1$ is at most three-dimensional.  From
the cohomology classes we already have, we can construct three: $[J_-
\theta]$, $[J_3 \theta]$, and $[J_+ \theta]$.

Let us summarize.  The dimension zero subspace of $\CA^0$ is generated
by $[1]$ and the dimension one subspace of $\CA^1$ is generated by
$[\theta]$, whereas the dimension one subspace of $\CA^0$ is generated
by $[J_-]$, $[J_3]$, $[J_+]$ and the dimension two subspace of $\CA^1$
is generated by $[J_- \theta]$, $[J_3 \theta]$, $[J_+ \theta]$.
Therefore, for the first two nontrivial dimensions, we find an
isomorphism $\CA^0 \iso \CA^1$ given by the map
\begin{equation}
  [\CO] \longmapsto [\CO\theta].
\end{equation}
It has been shown \cite{Malikov:1998dw} that this isomorphism persists
in higher dimensions.  This is as though $[1]$ and $[\theta]$ are
vacua of CFT --- both of them are annihilated by $\del_z$ --- and the
elements of $\CA^0$ are creation operators acting on these classes to
generate the rest of the $Q$-cohomology classes.  It is remarkable
that such a structure emerges despite the lack of conformal
invariance.  In order to understand where this structure comes from,
we must go beyond perturbation theory.

\section{Nonperturbative vanishing of chiral algebras}
\label{vanishing}

In the previous sections we have seen that quantum corrections deform
the classical chiral algebra, but perturbatively the deformation can
be understood within the framework of a cohomology theory on the
target space.  This is because in perturbation theory, one considers
fluctuations localized around constant maps.

Instantons are not quite like constant maps, but have a finite size in
the target space.  Their presence may therefore lead to deformations
of different kinds.  In this section, we will see a particularly
striking example: instantons annihilate all of the perturbative
$Q$-cohomology classes, making the chiral algebra trivial
nonperturbatively.  The existence of such a phenomenon was first
predicted by Witten \cite{Witten:2005px}, and subsequently confirmed
by Tan and the author \cite{Tan-Yagi-1}; see Arakawa and Malikov
\cite{Arakawa:2009cb} for a mathematical interpretation of Witten's
prediction.  Here we generalize the results of \cite{Tan-Yagi-1,
  Tan-Yagi-2, MR2415553} and establish a vanishing ``theorem'' for
chiral algebras.  We then explain how the vanishing of the chiral
algebra implies the spontaneous breaking of supersymmetry and the
absence of harmonic spinors on the loop space of the target space.

\subsection{$\boldsymbol{\CP^1}$ Model, with instantons}

The simplest example of a vanishing chiral algebra is provided by the
$\CP^1$ model which we studied in Section~\ref{PCA-CP1}.  As we saw
there, the perturbative chiral algebra of the $\CP^1$ model has the
structure of a Fock space: $[1]$ and $[\theta]$ play the role of
``ground states'', on which infinite towers of bosonic and fermionic
``excited states'' are constructed by acting with ``creation
operators'', namely bosonic $Q$-cohomology classes.  In view of this
suggestive structure, one may expect that instantons ``tunnel''
between the ``ground states'' and ``lift'' them out of the chiral
algebra.  This is indeed the case.

We now show that the perturbative $Q$-cohomology classes $[1]$ and
$[\theta]$ are annihilated together by instantons via the relation
\begin{equation}
  \label{Qt1}
  \{Q, \theta\} \propto 1 \, .
\end{equation}
This relation says that the equation $1 = 0$ holds in the
$Q$-cohomology.  Therefore the chiral algebra of the $\CP^1$ model
vanishes nonperturbatively.  A half of the perturbative $Q$-cohomology
classes are annihilated because their representatives become
$Q$-exact.  For classes $[\CO]$ in this category, we have
$\{Q,\CO\theta\} \propto \CO$.  Thus $[\CO]$ are annihilated together
with $[\CO\theta]$, and the latter should constitute the other half of
the perturbative chiral algebra.  This explains the observed Fock
space structure.

Since $c_1(\CP^1) = 2$, instantons of degree $k$ (which wrap the
target space $k$ times) violate R-charge by $2k$ and dimension by $k$.
Then the instanton corrections to the action of $Q$ on $\theta$ take
the form
\begin{equation}
  \{Q,\theta\} = \sum_{k=1}^\infty \Lambda^{2k} e^{-kt_0} \CO_k \, ,
\end{equation}
with $\CO_k$ being local operators of R-charge $2-2k$ and dimension
$(1-k,0)$.  Here, $t_0$ is the topological invariant $S_B$ evaluated
for instantons of degree one at the dynamical scale $\Lambda$.  There
are no local operators of negative dimension, hence $\CO_k = 0$ for $k
> 1$.  The remaining operator, $\CO_1$, is perturbatively $Q$-closed.
From the fact that for R-charge zero and dimension zero, $[1]$ is the
only perturbative $Q$-cohomology class and there are no perturbatively
$Q$-exact local operators, it follows $\CO_1 \propto 1$.  So if we can
show $\{Q,\theta\} \neq 0$, we establish the relation $\{Q, \theta\}
\propto 1$.

To see whether $\{Q,\theta\}$ is zero or not, we put it on a disk
(embedded in the worldsheet) and evaluate the path integral for
suitable boundary conditions.  For our purpose, we can compactify the
disk to a sphere and consider those boundary conditions that can be
represented by vertex operators at $\infty$.  We will therefore
compute the correlation function
\begin{equation}
  \label{QtO}
  \Bigvev{\CO(\infty) \oint \! d\zb \, G(\zb) \theta(0)}
\end{equation}
on $\Sigma = \CP^1$ for some local operators $\CO$.  We anticipate
that $\{Q,\theta\}$, expressed here as the contour integral of the
supercurrent $G$ around $\theta$, will be replaced by $1$ in the final
result.  To obtain a nonvanishing answer, then, we should take $\CO$
to be local operators of R-charge zero and dimension zero, i.e.,
functions on $X$.

The contributions to this correlation function should come from
instantons of degree one.  These are biholomorphic maps from $\Sigma =
\CP^1$ to $X = \CP^1$, whose moduli space $\CM_1 \iso \PGL_2(\C)$.  At
$\phi_0 \in \CM_1$, the number of $\psib_+$ and $\psi_+$ zero modes
are, respectively, given by the zeroth and first Hodge numbers of the
bundle $K_{\Sigma}^{1/2} \otimes \phi_0^*T_X \iso \CO_{\CP^1}(1)$.
Note that we have untwisted the theory before compactifying, because
the twisted $\CP^1$ model is anomalous on $\CP^1$.  Since
$h^0(\CO_{\CP^1}(1)) = 2$ and $h^1(\CO_{\CP^1}(1)) = 0$, there are two
$\psib_+$ zero modes and no $\psi_+$ zero modes.  These zero modes can
be absorbed by the $\psib_+$ fields in $G$ and $\theta$ without
bringing down interaction terms.  Then, up to the ratio of the bosonic
and fermionic determinants, the correlation function is computed to
leading order by dropping quantum fluctuations and integrating over
the fermion zero modes as well as the instantons:
\begin{equation}
  \label{intGtO}
  e^{-S_B(\phi_0)} \! \int \! d\CM_1 \, dc_0^1 \, dc_0^2 \,
  \CO(\infty)
  \oint \! d\zb \, G(\zb) \theta(0) \Bigr|_{\phi = \phi_0} \, .
\end{equation}
Here $d\CM_1$ is the measure on $\CM_1$, and $c_0^1$, $c_0^2$ are the
zero mode coefficients of the mode expansion of $\psib_+$.  We can
ignore subleading contributions.

Of course, we must evaluate the contour integral before dropping
quantum fluctuations.  (Without short distance singularities this
would vanish!)  It turns out that this seemingly straightforward task
is actually very tricky.  We are looking for an antiholomorphic single
pole $1/\zb$ in the OPE
\begin{equation}
  \label{Jt}
  G(\zb) \theta(0)
  = (g_{\phi\phib} \del_{\zb}\phi \psib_+)(\zb)
    \Bigl(-\frac{1}{2} R_{\phi\phib} \del_z\phi\psib_+\Bigr)(0) \, .
\end{equation}
It may appear that one can obtain such a pole by contracting
$\del_{\zb}\phi$ with $R_{\phi\phib}$.  However, this does not work
because the residue is just the classical action of $Q$ on $\theta$
and vanishes.  We must find additional antiholomorphic poles that
emerge nonperturbatively.

At this point, we recall that the fermionic fields take values in the
pullback of the tangent bundle of $X$ by the bosonic field.  Thus, the
eigenmodes in which they are expanded depend on the bosonic field,
which is itself subject to quantum fluctuations.  As a result, the
fermion modes --- even the zero modes --- can produce short distance
singularities when the bosonic field is present at the same location.

We can try to extract this bosonic dependence of the fermionic fields
as follows.  Consider a tubular neighborhood $\CN_1$ of $\CM_1$,
diffeomorphic to the normal bundle of $\CM_1$ in the space of maps
from $\Sigma$ to $X$.  Let $\{x^\alpha\}$ be local coordinates on
$\CM_1$ and parametrize the normal directions by coordinates
$\{y^\beta\}$ such that $y^\beta = 0$ on $\CM_1$.  For
$\phi(z,\zb;x,y) \in \CN_1$, we denote its projection to $\CM_1$ by
$\phi_0(z;x)$.  An instanton $\phi_0 \in \CM_1$ maps the points of
$\Sigma = \CP^1$ to the points of $X = \CP^1$ in a one-to-one manner,
so we can invert $\phi_0(z;x)$ to obtain $z(\phi_0;x)$ and write
$\phi(z, \zb; x, y) = \phi(\phi_0(z;x), \phib_0(\zb;x); x, y)$.
Computing $[iQ,\phib]$ with this last expression, we find
\begin{equation}
  \label{psib}
  \psib_+
  = \frac{\del\phib}{\del\phib_0} [iQ,\phib_0] + \dotsb
  = \frac{\del_\zb\phib}{\del_\zb\phib_0} \psib_{+0}(\phi_0) + \dotsb
    \, ,
\end{equation}
where $\psib_{+0}(\phi_0) = [iQ,\phib_0]$ is the zero mode part of
$\psib_+$ evaluated at $\phi_0$.  So we have extracted, partially, the
dependence of $\psib_+$ on the bosonic fluctuations.

In fact, the leading term of the formula \eqref{psib} is all we need.
The reason is that the fermion nonzero modes will be discarded in our
computation and, up to the nonzero modes and the equation of motion
for the bosonic field, the leading term represents the zero mode part
of $\psib_+$.  Indeed, it correctly reduces to $\psib_{+0}(\phi_0)$ at
$\phi = \phi_0$ and the equation of motion implies
\begin{equation}
  D_z\Bigl(\frac{\del_\zb\phib}{\del_\zb\phib_0} \psib_{+0}(\phi_0)\Bigr)
  = R^{\phib}{}_{\phib\phi\phib} \frac{\del_z\phib}{\del_\zb\phib_0}
    \psi_+ \psib_+ \psib_{+0}(\phi_0) \, ,
\end{equation}
which vanishes if the fermion nonzero modes are dropped since there
are no $\psi_+$ zero modes.

As desired, $\del_\zb\phi$ from $G$ can now be contracted with
$\del_\zb\phib$ in the $\psib_+$ field from $\theta$ to produce an
antiholomorphic double pole.  This gives
\begin{equation}
  \label{Gt}
  \oint \! d\zb \, G(\zb) \theta(0)
  =
  \Bigl(\frac{i}{2} R_{\phi\phib}
  \frac{\del_z\phi}{\del_\zb\phib_0}
  \del_\zb\psib_+ \psib_{+0}(\phi_0)\Bigr)(0) \, .
\end{equation}
The result may look strange, but will become more natural after the
$\psib_+$ zero modes are integrated out.

To proceed, we need to specify the path integral measure.  On
instantons, the action of $Q$ is realized as the superconformal
transformation $\zb \mapsto \zb + \epsilonb_-(c_0^1 + c_0^2 \zb)$.
Thus the zero mode part of $\psib_+$ can be expanded as
\begin{equation}
  \psib_{+0}(\phi_0)
  = c_0^1 \del_\zb\phib_0 + c_0^2 \zb\del_\zb\phib_0 \, .
\end{equation}
If we choose $d\CM_1$ to be conformally invariant, it is $Q$-invariant
up to terms involving $dc_0^1$ or $dc_0^2$.  Then the product $d\CM_1
\, dc_0^1 \, dc_0^2$ is a $Q$-invariant measure since $dc_0^1 \,
dc_0^2$ is $Q$-invariant by itself.  There is a unique conformally
invariant measure on $\CM_1$ up to a factor.  In terms of the points
$X_0$, $X_1$, $X_\infty \in X$ to which $0$, $1$, $\infty \in \Sigma$
are mapped by instantons, it is given by
\begin{equation}
  \label{dM1}
  d\CM_1
  = \frac{d^2\!X_0 \, d^2\!X_1 \, d^2\!X_\infty}
         {|X_0 - X_1|^2 |X_1 - X_\infty|^2 |X_\infty - X_0|^2} \, .
\end{equation}
The parametrization of $\CM_1$ by $X_0,$ $X_1$, $X_\infty$ provides a
compactification of $\CM_1$ to $(\CP^1)^3$, but $d\CM_1$ is singular
on the compactified moduli space.

Let us return to the integral \eqref{intGtO}.  After the integration
over the $\psib_+$ zero modes, the contour integral
\begin{equation}
  \label{GtRicci}
  \int \! dc_0^1 \, dc_0^2 \oint \! d\zb \, G(\zb)
  \theta(0)\Bigr|_{\phi = \phi_0}
  = \Bigl(\frac{i}{2} R_{\phi\phib} \del_z\phi_0 \del_\zb\phib_0\Bigr)(0) \, .
\end{equation}
This is the pullback of the Ricci form by instantons.  Using the
formula
\begin{equation}
  \del_z\phi_0(0)
  = \frac{(X_\infty - X_0)(X_0 - X_1)}{X_1 - X_\infty} \, ,
\end{equation}
what is left can be written as
\begin{equation}
  \label{XXX}
  \frac{i}{2} e^{-S_B(\phi_0)} \! \int_{\CP^1} d^2X_0 \, R_{\phi\phib}(X_0, \Xb_0)
  \int_{(\CP^1)^2}  d^2X_1 \, d^2X_\infty
  \frac{\CO(X_\infty, \Xb_\infty)}{|X_1 - X_\infty|^4} \, .
\end{equation}
The $X_0$-integral is the evaluation of $2\pi c_1(X)$ on $[X]$, which
gives $4\pi$.  The $X_1$-integral diverges, reflecting the
noncompactness of $\CM_1$.  One way to regularize it is to impose a
lower bound $l > 0$ on the distance between $X_1$ and $X_\infty$
measured with the target space metric.  For $l \ll 1$, this restricts
the domain of $X_1$ to the region where
$g_{\phi\phib}(X_\infty,\Xb_\infty) |X_1 - X_\infty|^2 \geq l^2$.  The
regularized integral is
\begin{equation}
  \int_{\CP^1} \frac{d^2X_1}{|X_1 - X_\infty|^4}
  = \frac{\pi}{l^2} g_{\phi\phib}(X_\infty, \Xb_\infty) \, .
\end{equation}
From this and the renormalization group equation \eqref{BREN} for the
$B$ field, we find that the integral \eqref{XXX} contains a factor
$(\Lambda/l\mu)^2 e^{-t_0}$.  We can take a limit such that $\mu \to
\infty$ and $l \to 0$ keeping $l\mu$ finite.  The end result is
\begin{equation}
  \Lambda^2 e^{-t_0}
  \int_{\CP^1} g_{\phi\phib} d^2X_\infty \,  \CO \, ,
\end{equation}
up to an overall numerical factor.

We see that the resulting $X_\infty$-integral is performed over $\CM_0
\iso X$ with respect to a natural volume form.  Therefore, the
one-instanton computation of the correlation function \eqref{QtO} has
reduced to the zero-instanton computation of the expected one-point
function:
\begin{equation}
  \bigvev{\CO(\infty) \{Q, \theta(0)\}}
  \propto \Lambda^2 e^{-t_0} \bigvev{\CO(\infty)} \, .
\end{equation}
This shows $\{Q, \theta\} \propto 1$.

\subsection{Vanishing ``theorem''}
\label{NV-VT}

The vanishing of the chiral algebra is not unique to the $\CP^1$
model.  It seems to be a feature shared by many $(0,2)$ models.  In
fact, we have the following vanishing ``theorem''.  Let $X$ be a
compact spin K\"ahler manifold with $p_1(X)/2 = 0$ and $c_1(X) > 0$.
Suppose that there is an embedding $\CP^1 \subset X$ with trivial
normal bundle.  Then, the chiral algebra of the $(0,2)$ model with
target space $X$ vanishes nonperturbatively in the absence of
left-moving fermions.

As before, we will establish the vanishing by demonstrating that
instantons annihilate the perturbative $Q$-cohomology classes $[1]$
and $[\theta]$ together.  Since $c_1(X) > 0$, every instanton violates
R-charge by $2k$ and dimension by $k$ for some integer $k > 0$.  Then
$[\theta]$ can only be annihilated by instantons with $k = 1$, in
which case it is paired with a class of R-charge zero and dimension
zero.  Such a class must be proportional to $[1]$ on the compact
complex manifold $X$.  Thus, it suffices to show $\{Q,\theta\} \neq
0$.

Consider the correlation function \eqref{QtO} again.  The leading
contributions to this function come from instantons that give
precisely two $\psib_+$ zero modes and no $\psi_+$ zero modes.  These
are embeddings $\CP^1 \hookrightarrow X$ whose normal bundle is
trivial.  (The normal bundle $N_{C/X}$ of a curve $C \subset X$ is a
holomorphic vector bundle over $C$ defined by the short exact sequence
$0 \to T_C \to T_X|_C \to N_{C/X} \to 0$.)  To see this, note that
given instantons wrapping a $\CP^1 \subset X$ once, there are already
the right amount of zero modes from the tangent direction.  So there
should be no additional zero modes from the normal directions, which
is the case if and only if the normal bundle of the $\CP^1$ is
trivial.

But if the normal bundle is trivial and the fermion zero modes come
only from the tangent direction, the contribution from each such
$\CP^1$ can be computed essentially in the same way as in the $\CP^1$
model; the result is a function on $X$ supported on the $\CP^1$.  By
assumption, there is at least one $\CP^1$ that makes a contribution,
hence the correlation function is nonzero.%
\footnote{In case the contributions from all the instantons add up to
  zero, we can insert delta functions supported on a particular
  $\CP^1$ at various points on $\Sigma$ so that the target space
  effectively becomes that $\CP^1$.  The correlation function is still
  nonzero.}
It follows that $\{Q, \theta\} \neq 0$ and the chiral algebra vanishes.

One may wonder how the existence of a single $\CP^1$ with trivial
normal bundle, whose contribution to $\{Q,\theta\}$ is confined on the
$\CP^1$ itself, can possibly tell something about the behavior of
$\{Q,\theta\}$ in the other region of $X$.  This point can be
understood if we consider deformations of the $\CP^1$.  Infinitesimal
deformations are given by holomorphic sections of the normal bundle.
Since the normal bundle is trivial, the $\CP^1$ in question can be
moved in every direction in the target space.  Then the family of
$\CP^1$s generated by deformations sweep out the whole target space,
and their contributions can add up to a nonzero constant.

\subsection{Flag manifold model}

An example in which the above deformation argument is beautifully
demonstrated is the flag manifold $G/B$ of a complex simple Lie group
$G$ with Borel subgroup $B$ \cite{Tan-Yagi-2, MR2415553}.  For $X =
G/B$, we have $p_1(X)/2 = 0$ and $c_1(X) = 2(x_1 + \dotsb + x_r)$ with
$r = \rank G$.  Thus, the $G/B$ model is well defined and has the
R-symmetry broken to $\Z_2$ nonperturbatively.  The simplest case is
when $G = \SL_2$, for which $G/B \iso \CP^1$.

What makes the $G/B$ model interesting is that its perturbative chiral
algebra contains currents generating the affine Lie algebra $\afg$ of
critical level \cite{Malikov:1998dw, MR1042449, MR2290768}.  The
critical level makes an appearance here because the Sugawara
construction must fail; otherwise, it would contradict the fact that
the chiral algebra lacks an energy-momentum tensor when $c_1(X) \neq
0$.  Nonperturbatively, these currents disappear from the chiral
algebra along with everything else.

Pick a $\CP^1 \subset X$ and call it $C$.  Choose linearly independent
normal vectors $V_1$, $\dotsc$, $V_d \in N_{C/X}|_{g_0}$ at a point
$g_0 \in C$.  Under $g_0 \mapsto gg_0 \in C$, these are mapped to
$g_*V_1$, $\dotsc$, $g_*V_d \in N_{C/X}|_{gg_0}$, which are again
linearly independent.  Varying $g$, we obtain a global frame of
$N_{C/X}$.  Therefore $N_{C/X}$ is trivial, and the chiral algebra
vanishes.  In this example, the $G$-action generates a family of
$\CP^1$s with trivial normal bundle that covers the target space.

\subsection{Supersymmetry breaking and the geometry of loop spaces}

The vanishing of the chiral algebra of a $(0,2)$ model has important
implications for the dynamics of the theory and the geometry of the
loop space $\CL X$ of the target space: supersymmetry is spontaneously
broken and there are no harmonic spinors on $\CL X$.  These
conclusions are obtained by studying the $Q$-cohomology of states
rather than operators.

Since $Q$ has R-charge one and satisfies $Q^2 = 0$, one can consider
the $Q$-cohomology graded by the R-charge in the Hilbert space of
states.  This is naturally a module over the chiral algebra: on
elements $[\ket{\Psi}]$ of the $Q$-cohomology of states, $[\CO] \in
\CA$ acts by
\begin{equation}
  [\CO] \cdot [\ket{\Psi}] = [\CO\ket{\Psi}] \, .
\end{equation}
As a graded vector space, it is isomorphic to the space of
supersymmetric states, the kernel of $\{Q,Q^\dagger\} \propto H - P$
which is generally infinite-dimensional since $P$ is not bounded.  If
$c_1(X) = 0$, it is further isomorphic to the $Q$-cohomology of local
operators by the state-operator correspondence.

To unravel the geometric meaning of the $Q$-cohomology of states, take
$\Sigma$ to be a cylinder $S^1 \times \R$ with coordinates $(\sigma,
\tau)$ and regard $\tau$ as time.  The theory may now be viewed as
supersymmetric quantum mechanics on $\CL X$, so let us canonically
quantize it and see what we get.  The fermionic fields are quantized
(for $z = \sigma + i\tau$ and with respect to a local orthonormal
frame) to obey
\begin{equation}
  \label{CAR}
  \{\psi_+^a(\sigma, \tau), \psib_+^\bb(\sigma', \tau)\}
  = \delta^{ab} \delta(\sigma - \sigma') \, .
\end{equation}
This is a loop space version of the Clifford algebra, with the
continuous index $\sigma$ parametrizing the direction along the loop.
States are thus spinors on $\CL X$.  The supercharge is quantized as
\begin{equation}
  \label{Qquant}
  Q = -i\int \! d\sigma \, \psib_+^\ib 
      \Bigl(\frac{D}{\delta\phi^\ib}
            - B_{\ib j} \del_\sigma\phi^j
            - B_{\ib\jb} \del_\sigma\phi^\jb\Bigr) \, ,
\end{equation}
where $D/\delta\phi$ is the covariant functional derivative on $\CL
X$.  From this expression, we see that $Q$ is almost a half of the
Dirac operator on $\CL X$.  But not quite, since it has extra pieces
coupled to $\del_\sigma$.

We can eliminate these extra pieces without changing the
$Q$-cohomology.  To do this, we define a functional $\CA_B\colon \CL X
\to \C$ as follows.%
\footnote{This functional is not single valued if the cohomology class
  $[B]$ does not vanish on some two-cycles.  However, one can always
  go to a covering space of $\CL X$ in which it becomes single valued,
  and define the theory in this space.  See \cite{MR2003030} for more
  discussion on this point.}
First, we pick a base loop in each connected component of $\CL X$.
Then, given $\phi \in \CL X$, we choose a homotopy $\phih\colon [0,1]
\times S^1 \to X$ from the base loop of the relevant component to
$\phi$.  Finally, we set
\begin{equation}
  \label{h}
  \CA_B(\phi) = \int_{[0,1] \times S^1} \phih^* B \, .
\end{equation}
Under a variation of the end point $\phi \to \phi + \delta\phi$, this
functional changes by
\begin{equation}
  \label{deltah}
  \delta \CA_B
  = \int \! d\sigma \bigl(\delta\phi^i
     (B_{ij} \del_\sigma\phi^j
      + B_{i\jb}\del_\sigma\phi^\jb)
     - \delta\phi^\ib
       (B_{\ib j}\del_\sigma\phi^j
        + B_{\ib\jb} \del_\sigma\phi^\jb)\bigr) \, .
\end{equation}
Writing $Q_0$ for a half of the Dirac operator obtained by dropping
the extra pieces from $Q$, we have
\begin{equation}
  \label{Qh}
  Q = e^{-\CA_B} Q_0 e^{\CA_B} \, .
\end{equation}
Therefore, the $Q$-cohomology of states is isomorphic to the
$Q_0$-cohomology, which is the cohomology of the spinor bundle over
$\CL X$.

Now suppose that the chiral algebra is trivial: $[1] = 0$.  Then
\begin{equation}
  [\ket{\Psi}] = [1] \cdot [\ket{\Psi}] = 0
\end{equation}
for any $Q$-cohomology classes $[\ket{\Psi}]$, so the $Q$-cohomology
of states is also trivial.  Since supersymmetric states are in
one-to-one correspondence with the $Q$-cohomology classes, there are
none.  In other words, supersymmetry is spontaneously broken.
Furthermore, we just saw that the $Q$-cohomology is nothing but the
cohomology of the spinor bundle over $\CL X$.  Hence, there are no
harmonic spinors on $\CL X$ either.

\section*{Acknoweldgments}

This work is based on the author's Ph.D. thesis.  First and foremost,
I would like to thank the Department of Physics and Astronomy at
Rutgers University for providing a stimulating environment during my
graduate studies.  I am also grateful to the Center for Frontier
Science at Chiba University for their support while this paper was
written.  I am indebted to my advisor Gregory Moore, whose deep
understanding and original thinking have always been a great
inspiration to me.  Last but not least, my gratitude goes to
Meng-Chwan Tan for being such a wonderful colleague and friend.

\providecommand{\href}[2]{#2}

\end{document}